\documentclass[11pt,english]{article}
\usepackage{fullpage}

\usepackage[T1]{fontenc}
\usepackage[latin9]{inputenc}
\usepackage{amsmath}
\usepackage{amssymb}
\usepackage{mathtools}
\usepackage{algorithmicx}
\usepackage{amsthm}
\usepackage{thm-restate}
\usepackage{blindtext}
\usepackage{comment}
\usepackage{enumitem}
\usepackage{xr}  
\usepackage[dvipsnames]{xcolor}
\usepackage{natbib}
\usepackage[english]{babel} 
\usepackage{tikz}

\newtheorem{proposition}{Proposition}[section]
\newtheorem{lemma}[proposition]{Lemma}
\newtheorem{example}[proposition]{Example}

\newtheorem{theorem}[proposition]{Theorem}
\newtheorem{definition}[proposition]{Definition}
\newtheorem{corollary}[proposition]{Corollary}
\newtheorem{observation}[proposition]{Observation}
\newtheorem{maintheorem}{Main Theorem}
\newtheorem*{theorem*}{Theorem}
\newtheorem*{definition*}{Definition}

\makeatletter
\@ifundefined{date}{}{\date{}}
\makeatother
\usepackage{fullpage}
\usepackage{graphicx}
\usepackage{algorithm} 

\graphicspath{{figures/}}

\usepackage[]{color-edits}
\usepackage{url}
\usepackage[noend]{algpseudocode} 
    
\addauthor{MF}{red}

\addauthor{LY}{NavyBlue}

\addauthor{XX}{Green}

\newcommand{\reals}{\mathbb{R}}

\newcommand{\Sa}{S_{\alpha}}
\newcommand{\Sae}{S_{\alpha-\epsilon}}

\DeclareMathOperator*{\argmax}{arg\,max}

\newcommand{\SPA}{SPA}

\newcommand{\GS}{\textsf{GS}}
\newcommand{\gs}{\textsf{gross substitutes}}

\newcommand{\Ultra}{\textsf{Ultra}}

\newcommand{\wl}{\textsf{well-layered}}

\newcommand{\WL}{\textsf{WL}}
\newcommand{\wwl}{\textsf{weakly well-layered}}
\newcommand{\Wwl}{\textsf{Weakly well-layered}}
\newcommand{\WWL}{\textsf{WWL}}

\newcommand{\GreedyGSVal}{\textsc{GreedyGS1}}
\newcommand{\GreedyGSContract}{\textsc{GreedyGS2}}
\newcommand{\GreedyGSSPA}{\textsc{GreedyGS-SPA}}
\newcommand{\AltGreedyGSSPA}{\textsc{Alt\GreedyGSSPA}}

\newcommand{\GreedyUltraVal}{\textsc{GreedyUltra1}}
\newcommand{\GreedyUltraContract}{\textsc{GreedyUltra2}}
\newcommand{\GreedyUltraSPA}{\textsc{GreedyUltraSPA}}
\newcommand{\AltGreedyUltraSPA}{\textsc{Alt\GreedyUltraSPA}}

\newcommand{\GreedyUpToT}{\textsc{GreedyUpToT}}

\newcommand{\GreedyWWL}{\textsc{Greedy-WWL-SYM}}

\newcommand{\DemandForSPA}{\textsc{GreedyForSPA}}

\newcommand{\setfunc}{2^A \to \reals_{\geq 0}}
\newcommand{\pricevec}{\reals^{n}_{\geq 0}}
\newcommand{\subA}{\subseteq A }

\usepackage{pifont} 

\usepackage{authblk}

\begin{document}

\author{Michal Feldman}
\author{Liat Yashin}
\affil{Blavatnik School of Computer Science, Tel Aviv University}
\title{Ultra Efficient Contracts: \\Pushing the Boundaries of Tractable Contract Design\thanks{
This project has been partially funded by the European Research Council (ERC) under the European Union's Horizon 2020 program (grant agreement No.~866132), by the European Union's Horizon Europe Program (grant agreement No.~101170373), by an Amazon Research Award, by the Israel Science Foundation Breakthrough Program (grant No.~2600/24), and by the NSF-BSF (grant number 2020788)
}
}
\maketitle
\thispagestyle{empty} 

\begin{abstract}

We study the optimal contract problem in the \emph{combinatorial actions} framework of D\"utting et al.~[FOCS'21], where a principal delegates a project to an agent who chooses a subset of hidden, costly actions, and the resulting reward is given by a monotone set function over the actions.
The principal offers a contract that specifies the fraction of the reward the agent receives, and the goal is to compute a contract that maximizes the principal's expected utility.
%
Prior work established polynomial-time algorithms for \emph{gross substitutes} rewards, while showing NP-hardness for general submodular rewards; subsequent work extended tractability to \emph{supermodular} rewards, demonstrating that tractable cases exist in both the substitutes and complements regimes. This left open the precise boundary of tractability for the optimal contract problem.


Our main result is a polynomial-time algorithm for the optimal contract problem under \Ultra\ rewards, a class that strictly contains gross substitutes but is not confined to subadditive rewards, thereby bridging the substitutes and complements regimes.
%
%
We further extend our results beyond additive costs, establishing a polynomial-time algorithm for \Ultra\ rewards and cost functions that are the sum of additive and symmetric functions. To the best of our knowledge, this is the first application of \Ultra\ functions in a prominent economic setting.
\end{abstract}


\begin{titlepage}

\maketitle

\end{titlepage}

\section{Introduction}


As digital markets for services scale and diversify, the problem of designing incentive schemes
that are both computationally tractable and guarantee desirable outcomes has become a central
focus of modern algorithmic game theory. For comprehensive surveys of recent work in this area, see~\cite{DFT24survey,Feldman26}. A prominent abstraction in this line of work is the
model of \emph{combinatorial contracts}, introduced by D\"utting et al.~\cite{DEFK21}.
In this model, a principal delegates a project to an agent who may select any subset
$S \subseteq A$ of costly actions. Each action $a \in A$ incurs a cost $c(a)$, and the chosen set of actions is hidden from the principal.
The project has a binary outcome: it succeeds with probability $f(S)$ and yields a
normalized reward of $1$ to the principal, or fails and yields $0$. The function
$f : 2^{A} \to [0,1]$ is therefore both the success probability and the expected reward, and is assumed to be monotone.

The principal incentivizes the agent via a linear contract $\alpha \in [0,1]$, under which the agent
receives an $\alpha$-fraction of the realized reward. Given a contract, the agent chooses a set
$S$ maximizing her expected payment minus cost, where any such set is denoted an  \emph{agent's best response}. Anticipating this best response, the principal's
goal is to compute the $\alpha$ that maximizes her own expected utility, given by the expected reward minus expected payment.


A central algorithmic question in this framework is to characterize which classes of reward
functions admit poly-time computation of an optimal contract.
Two such results have been established, on both sides of the substitutes-complements divide.

\paragraph{Gross substitutes rewards.}
A key result of~\cite{DEFK21} shows that when the reward function $f$ belongs to the class of
\emph{gross substitutes} (GS) --- a strict subclass of submodular functions --- the optimal contract can
be computed in polynomial time. This positive result stands in sharp contrast to their accompanying
hardness result, showing that the problem becomes NP-hard for general submodular reward
functions. 




\paragraph{Supermodular rewards.}
Subsequent work by \cite{DFG24combinatorial,DDPP24} demonstrated that tractability is not confined to substitutes.
In particular, polynomial-time algorithms were established for \emph{supermodular} reward functions --- a class that is incomparable with GS and lies outside the complement-free hierarchy of~\cite{LLN06}.

These results lead to the following natural question:


\begin{quote}
\emph{Which classes of reward functions admit polynomial-time algorithms for computing
the optimal contract?}
\end{quote}


To gain further intuition for the optimal contract problem, we revisit the core observations and techniques that underlie existing tractable cases.

At a high level, an efficient algorithm for the optimal contract problem relies on two key ingredients: (1) the ability to compute the agent's best response efficiently, and (2) the fact that the agent's best response changes only at polynomially-many values of the contract parameter~$\alpha$, known as \emph{critical values}.
Specifically, it was shown by \cite{DFG24combinatorial} that, combined with the Eisner-Severance algorithm \cite{EisnerSeverance1976}, one can transverse the ``upper envelope" of the agent's utility function (as a function of the contract parameter $\alpha$) to find the optimal contract.

Regarding the first ingredient, a central observation of~\cite{DEFK21} is that, for any fixed contract parameter~$\alpha$, the agent's best response problem can be formulated as a \emph{demand query} for the reward function~$f$. In this interpretation, actions are viewed as goods in a market, and $f$ plays the role of a valuation function over bundles of goods. Formally, given a valuation $f$ and a price vector
$p=(p_1,\ldots,p_n)\in\mathbb{R}^n_{\ge 0}$, a demand query asks for a set
$S\subseteq A$ maximizing the utility $u(S)=f(S)-\sum_{a\in S}p_a$.
In the contract problem under a linear contract~$\alpha$, the agent chooses a set~$S$ maximizing
$\alpha\cdot f(S)-\sum_{a\in S}c(a)$, which is equivalent to a demand query with prices $p_a=c(a)/\alpha$. Consequently, the ability to compute demand queries efficiently for the reward function~$f$ emerges as a central algorithmic primitive for solving the optimal contract problem.

For \gs\ (\GS) reward functions, it is well known that demand queries can be computed in polynomial time via a greedy algorithm that repeatedly adds an element maximizing marginal utility and terminates once this marginal becomes negative (see, e.g., \cite{PaesLeme17}). The main technical contribution of~\cite{DEFK21} was to complement this fact by showing that, for \GS\ rewards, there are only polynomially many critical values of~$\alpha$, yielding a polynomial-time algorithm for the optimal contract problem.

\paragraph{\Ultra\ rewards.}
In this paper, we consider a third class of reward functions known as \Ultra\ rewards. This class was introduced by~\cite{DT95well} under the name \wl\ and later re-formalized by~\cite{Leh17}. Like \GS\ functions, \Ultra\ functions admit a characterization via a greedy process, though with important differences. The formal definition appears in Definition~\ref{def:ultra-new}; we provide an informal description for intuition.

Fix a price vector~$p$. Consider a greedy procedure that repeatedly adds an element maximizing the marginal utility, but, unlike the \GS\ greedy algorithm, continues until all elements have been considered, even if some marginal contributions are negative. This process generates a nested sequence of sets $\emptyset = S_0 \subseteq S_1 \subseteq \cdots \subseteq S_n$. A function is \Ultra\ if, for every index $i\in[n]$, the set~$S_i$ maximizes utility (with respect to prices~$p$) among all sets of cardinality~$i$.
Given this characterization, answering a demand query for \Ultra\ functions reduces to selecting the set $S_i$ with maximum utility (with respect to prices $p$) over all $i\in [n]$.



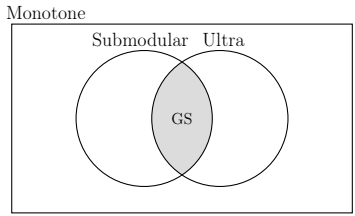
\begin{figure}[h]
    \centering
    \begin{minipage}{0.5\linewidth}
    \centering
    \scalebox{0.5}{
    \begin{tikzpicture}
        \definecolor{lightgray}{RGB}{220,220,220}
        
        \draw[thick] (-4.5,-2.5) rectangle (4.5,2.5);
        \node at (-3.6,2.8) {\Large Monotone};
        
        \begin{scope}
            \clip (-1,0) circle (1.8);
            \fill[lightgray] (1,0) circle (1.8);
        \end{scope}

        \draw[thick] (-1,0) circle (1.8);
        \draw[thick] (1,0) circle (1.8);
        
        \node at (-1.1,2.1) {\Large Submodular};
        \node at (1.1,2.1) {\Large Ultra};
        
        \node at (0,0) {\large GS};
    \end{tikzpicture}}
    
\end{minipage}
\hfill
\begin{minipage}{0.45\linewidth}
\caption{The relationship between monotone submodular, \Ultra, and \GS\ functions. \GS\ is precisely the intersection of submodular and \Ultra\ set functions.}
\label{fig:class-relations}
\end{minipage}
\end{figure}

Unlike \GS, \Ultra\ functions are not necessarily submodular and, in fact, extend strictly beyond the complement-free hierarchy of~\cite{LLN06}, including functions that are not subadditive.
As such, the \Ultra\ class naturally spans both substitutes and complements, allowing for controlled forms of complementarities while still retaining significant structure.
Nevertheless, the similarity between the greedy characterizations of \GS\ and \Ultra\ is not coincidental; rather, the two classes are closely related.
In particular, for monotone set functions, \GS\ is exactly the intersection of the submodular and \Ultra\ classes, as illustrated in Figure~\ref{fig:class-relations}.
Since prior work has established tractability results for the optimal contract problem on both sides of the substitutes-complements divide, the \Ultra\ class is of particular interest, as it bridges these previously separate regimes.

In this paper, we study the optimal contract problem for \Ultra\ reward functions.


%


\subsection{Our Contribution}

This paper advances the tractability frontier of combinatorial contract design in several ways.

First, we show that Ultra reward functions admit a polynomial-time algorithm for computing the optimal contract under additive costs. 


\begin{maintheorem}
[Poly-Time Algorithm for Ultra Rewards; Theorem 3.1]
Let $f$ be a monotone Ultra reward function and let $c$ be an additive cost function. Then the optimal contract problem can be solved in polynomial time using value queries.
\end{maintheorem}

Our proof approach follows the framework introduced by \cite{DEFK21}. In particular, we show how to compute an agent's best response in polynomial time, and provide a polynomial bound on the number of critical values for \Ultra\ rewards.

To compute an agent's best response, we use the reduction from the agent's best response to a contract $\alpha$ to a demand query with prices $c(a)/\alpha$. By the \wl\ property of \Ultra\ valuations, they admit a greedy demand oracle. We adapt this oracle to the contract setting by adding a tie-breaking rule that is aligned with
the principal's utility.

Our main technical contribution is proving a polynomial upper bound on the number of critical values in settings with \Ultra\ rewards and additive costs.
The analysis exploits the \wl\ structure of \Ultra\ functions to characterize how the agent's best response changes as the payment parameter $\alpha$ increases. 
In the spirit of \cite{DEFK21}, we then combine this characterization with a genericity perturbation and a potential-function argument to show that the total number of critical values is at most $O(n^2)$.

\medskip

Second, \cite{DEFK21} left open whether the \GS\ result extends to richer cost structures than additive. 
We shed light on this question by studying cost functions that decompose into the sum of an additive component and a symmetric component. 
This class strictly extends both purely additive and purely symmetric cost functions. 
Moreover, our analysis applies to the broader class of \Ultra\ reward functions, going beyond the \GS\ setting considered in \cite{DEFK21}. 

\begin{theorem*}[Beyond Additive Costs; Theorem~4.2]
Let $f$ be an \Ultra\ reward function, and let $c$ be a cost function that decomposes into the sum of an additive component and a monotone symmetric component. Then the optimal contract can be computed in polynomial time.
\end{theorem*}


To establish this result, we design a new best response algorithm for the setting of \Ultra\ rewards and costs that decompose into the sum of an additive component and a symmetric component. The algorithm exploits the closure of the class of \Ultra\ functions under addition with symmetric functions.

Bounding the number of critical values requires a more delicate analysis.
We introduce an agent's best response 
algorithm that is used solely for analytical purposes.
Our approach is based on a family of \emph{truncated reward functions}: for each $i$, the $i$-truncated reward function assigns to any set $S$ the maximum reward achievable by a subset of $S$ of size at most $i$.
We design an algorithm that computes the agent's best response with respect to these truncated reward functions.
Crucially, for every $i \in [n]$, the agent's best response under the $i$-truncated reward function coincides with the agent's optimal choice among all sets of cardinality at most $i$ under the original reward function.
Leveraging this equivalence together with the \wl\ property, we show how best response computations for truncated reward functions can be combined to recover the agent's best response for the original reward function.
Finally, we bound the number of critical values in the truncated setting, which in turn yields a bound on the number of critical values in the original problem.

\medskip

Finally, we establish tractability for a broader class of reward functions,  \emph{weakly well-layered},  \cite{efx-wwl} when costs are symmetric. \Wwl\ is a strict superclass of \Ultra. 

\begin{theorem*}[\Wwl\ Rewards; Appendix~\ref{sec:appendix-wwl-1} Proposition~\ref{prop:optimal-contract-wwl-sym}]
Let $f$ be a \wwl\ reward function and let $c$ be a symmetric cost function. Then the optimal contract can be computed in polynomial time.
\end{theorem*}

Interestingly, the \wwl\ class includes budget-additive rewards, for which the optimal contract problem is known to be NP-hard under additive costs \cite{DEFK21}.

\subsection{Additional Related Work}
\label{sec:related}

\paragraph{Algorithmic contract design.}
        The combinatorial-action model of contracts was introduced by \cite{DEFK21}, who provided the first algorithmic result for Gross Substitutes (\GS) reward functions and proved NP-hardness for general submodular rewards. 
        Subsequent work established an FPTAS for arbitrary monotone reward functions in \cite{DEFK25,DFG24combinatorial}.
        For supermodular rewards, tractability was later shown in \cite{DDPP24,DFG24combinatorial}, while hardness of approximation for submodular rewards under value queries was demonstrated in \cite{ezra2023Inapproximability}.
        In addition, for budgeted combinatorial contracts settings,
        \cite{FGPS26Budgeted} establishes strong impossibility results and identifies GS as a tractability frontier in budgeted environments.
        In parallel, \cite{DFTR26Information} establish hardness results under demand queries as well. 
    
        Several extensions of the core model have been explored. 
        Dynamic or sequential variants appear in \cite{ezra2024sequential,HSS25}. 
        Multi-agent and multi-action generalizations were studied in \cite{BFN06agency,BFNW12,DEFK25,CBCG24}. 
        Other lines of work address robust or learning-based contract design, including \cite{Car15,WC22,DT22,Kam23,PT24,CMG21,GSW21,GSWZ23,AG23,BTCXZ24,DFPS25pseudodim}.

\paragraph{\Ultra\ set functions.}
    \Ultra\ valuations were first introduced by~\cite{DT95well}, though under the name \wl, and were later re-formalized by~\cite{Leh17}. 
    \Ultra\ functions play a central role in discrete convex analysis. 
    \GS\ valuations are exactly the monotone submodular Ultra valuations \cite{DT95greedy}. 
    While \GS\ marks the boundary for guaranteed existence of Walrasian equilibria \cite{KJC82,GS99}, more general Ultra valuations need not admit an equilibrium. Whether classic welfare-maximization algorithms extend from \GS\ to Ultra remains an open question.

\section{Model and Preliminaries}

\subsection{Combinatorial Contracts Model}

\paragraph{The combinatorial-action contract model.}
We consider a setting where the principal delegates the execution of a project to a single agent. The project has a binary outcome: it either succeeds, yielding a reward for the principal, or fails, yielding nothing.
The agent may choose to perform any subset of actions from a given set $A = \{1, 2, \ldots, n\}$.

At the core of our model are two set functions: a cost function $c:\setfunc$,  which assigns a cost to each subset of actions (incurred by the agent),
and a reward function $f:2^A \to [0,1]$, which represents the probability of success given a chosen subset.

Without loss of generality, we normalize the reward for success to 1. Accordingly, $f$ also corresponds to the expected reward, and we refer to it as the reward function. 

We assume that $f(\emptyset) = 0$ and $c(\emptyset) = 0$, and that both $f$ and $c$ are monotone, which means that for any $S' \subseteq S$, it holds that $f(S') \leq f(S)$ and $c(S') \leq c(S)$.

\paragraph{Contract design.}
The principal incentivizes the agent to exert effort through a payment scheme, referred to as a {\em contract}.
Crucially, the principal cannot observe which actions the agent takes, but only the final outcome.
As a result, the contract is a payment function that specifies the agent's compensation in case of success. 
In this binary-outcome model, it is without loss of generality to restrict attention to a {\em linear} contract, 
defined by a single parameter $\alpha \in [0,1]$, specifying the fraction of the reward 
paid to the agent upon success.

This interaction forms a two-step game. In the first step, the principal offers the agent a contract $\alpha \in [0,1]$.
In response, the agent chooses a subset of actions 
to take. Finally, the principal receives their reward and pays the agent according to the contract $\alpha$. As standard, we impose a {\em limited liability} constraint, meaning that the payments can only flow from the principal to the agent.

\paragraph{Utilities and best responses.}
Both participants aim to maximize their respective utilities. The agent's utility is $u_a(\alpha,S) = \alpha f(S) -c(S)$. The {\em agent's best response} to a given contract $\alpha$ is any set that maximizes utility, with ties broken in favor of higher rewards.

Formally, let $D_{f,c}(\alpha) = \argmax_S u_a(\alpha, S)$ denote the collection of sets that maximize the agent's utility.
Among these, let $D^*_{f,c}(\alpha) = \argmax_{S\in D_{f,c}(\alpha)} f(S) \subseteq D_{f,c}(\alpha)$ be the sub-collection of sets with the highest reward value $f$.
Any set $\Sa \in D^*_{f,c}(\alpha)$ is considered an agent's best response.

The utility of the principal is given by $u_p(\alpha) = (1-\alpha) f(S_{\alpha})$.
Note that the agent's tie-breaking rule in favor of higher $f$ values effectively breaks the ties in favor of the principal's utility.

We also define the agent's {\em best response of size $i$} to be any set that maximizes the agent's utility among all sets of size $i$. The collection of all such sets is denoted by $D^i_{f,c}(\alpha) = \argmax_{S,|S|=i} u_a(\alpha,S)$.

Similarly, the agent's {\em best response of size at most $i$} is any set that maximizes the agent's utility among all sets of size {\em at most} $i$, and its collection is denoted by $D^{\leq i}_{f,c}(\alpha) = \argmax_{S,|S| \leq i} u_a(\alpha,S)$.

\paragraph{Value and Demand Oracles.}
In combinatorial settings, the reward function $f$ has exponential description size, and therefore is accessed using two standard oracle types: (i) a \emph{value oracle}, which evaluates $f(S)$ on a given set $S \subseteq A$, and (ii) a \emph{demand oracle}, which, given price vector $p \in \pricevec$, returns a set $S \subseteq A$ maximizing $f(S) - \sum_{a\in S} p_a$. 
A central observation of~\cite{DEFK21} is that the agent's best response under a linear contract $\alpha$ can be recast as a demand query: the agent selects a set $S$ maximizing $\alpha f(S) - \sum_{a\in S} c(a)$, which is equivalent to maximizing $f(S) - \sum_{a\in S} p_a$ with prices $p_a = c(a)/\alpha$. Thus, a demand oracle for $f$ can be used to compute the agent's best response. In general, demand oracles are stronger than value oracles, but in some special cases they can be efficiently implemented using value queries.



\subsection{Classes of Set Functions}
\label{sec:set-functions}

In this section we introduce different properties of set functions. 
Throughout this paper, we focus on {\em monotone} set functions, namely, $f(S) \leq f(T)$ for any $S \subseteq T$.

For presentation simplicity, we use the following shorthand operations on sets:
for any set $S\subA $, and element $t\notin S$, we denote the set $S \cup\{t\}$ by $S+t$. 
Similarly, for $s\in S$, we denote the set $S \setminus\{s\}$ by $S-s$, and denote the set $(S\setminus\{s\})\cup \{t\}$ by $S-s+t$.
In a similar way, for two sets $S,T\subseteq A$, we use $S-T$ to denote $S\setminus T$. 
For any set function $f:\setfunc$, the marginal value of an element $a \in A \setminus S$ given a set $S$ is $f(a \mid S) = f(S + a) - f(S)$. Lastly, for any $a\in A$, we use $f(a)$ as shorthand for $f(\{a\})$.

We first present several well-known set functions.

\begin{definition}
A set function $f:\setfunc$ is:
\begin{itemize}[itemsep=1pt]
  \item \emph{additive}
  if for every $S\subA$, 
  $f(S) = \sum_{a\in S} {f(a)}$.
  \item \emph{submodular}
  if for every two sets $S \subseteq T$
  and every $a \notin T$, $f(a \mid S) \geq f(a \mid T)$.
  \item \emph{gross substitutes (\GS)} if for any two price vectors 
  $p, q \in \pricevec$, with $p \leq q$ (pointwise) and any set $S \in D_f(p)$, there exists a set $T \in D_f(q)$ 
  such that $T \supseteq \{i\in S \mid p_i=q_i\}$.
  \item \emph{symmetric} if its value depends only on the cardinality of the set.
  That is, for every $S,T$ with $|S|=|T|$, $f(S)=f(T)$.
\end{itemize}
\end{definition}







We next define the class of \Ultra\ set functions. This class admits several equivalent characterizations; we begin with a definition based on the \emph{well-layered} property.

\begin{definition}[\wl\ (\WL)]\cite{DT95well}
\label{def:well-layered}
Let $f:\setfunc$ be a set function and $p\in\pricevec$ a price vector. 
Consider the sequence of sets $S_0 \subseteq S_1 \subseteq \ldots \subseteq S_n$, with $S_0=\emptyset$, constructed by the following greedy procedure:

For $i=1,\ldots,n$, let $S_i = S_{i-1} + x_i$, where
$x_i \in \argmax_{x\notin S_{i-1}} \Big\{f(x \mid S_{i-1})-p_x\Big\}$, with ties broken arbitrarily.

$f$ is said to be \wl\ (\WL) if for every price vector $p$, for every $i\in[n]$, it holds that
\[
S_i \in \argmax_{S,|S|=i} \Big\{f(S)-\sum_{x\in S}{p_x}\Big\}.
\]
\end{definition}

We next present two additional properties of set functions.

\begin{definition}[Exchange and Triplet properties]
\label{def:exchange-triplet}
Let $f:\setfunc$ be a set function. 
\begin{itemize}[itemsep=1pt]
\item
    $f$ satisfies the {\em exchange} property \cite{Leh17} if for any $S,T \subseteq A$ such that $ |S|  \leq  |T| $ and any $x \in S - T$, there exists some $y \in T - S$ such that $f(S) + f(T) \leq f(S - x + y) + f(T - y + x)$.

\item
    $f$ satisfies the {\em triplet property} \cite{reijnierse2002verifying} if for any $S\subA$ and any distinct triplet of $i,j,k\notin S$: $f(i+j \mid S) + f(k \mid S) \leq
    \max \Big\{  f(i \mid S) + f(j+k \mid S), 
    f(j \mid S) + f(i+k \mid S) \Big\}$.
\end{itemize}
\end{definition}

    


The following theorem shows that all three properties are equivalent.

\begin{theorem}\cite{Leh17}
\label{thm:ultra}
    The \wl\, exchange, and triplet properties are all equivalent.
\end{theorem}


A set function satisfying any of these properties is called \Ultra.

\begin{definition}[\Ultra]\cite{Leh17} \label{def:ultra-new}
    A set function $f$ is \Ultra\ if and only if it satisfies any of the properties \wl, exchange, or triplet.
\end{definition}

Interestingly, an alternative characterization of \GS\ functions is the following:

\begin{theorem} \cite{DT95greedy}
\label{thm:gs-is-wl}
    A set function $f$ is \GS\ if and only if: 
    \begin{itemize}[itemsep=1pt]
        \item $f$ is monotone submodular, and
        \item $f$ satisfies the \wl\ property. 
    \end{itemize}
\end{theorem}
Together with Theorem~\ref{thm:ultra}, it implies that for monotone set functions, \GS\ is precisely the intersection of submodular and \Ultra, see Figure~\ref{fig:class-relations}.
\paragraph{Remark:}
While \Ultra\ set functions are not necessarily monotone, in this work we focus on monotone set functions, and therefore restrict attention to monotone \Ultra\ set function.

\subsection{Demand Queries and Best Response} \label{sec:demand}

\paragraph {\GS\ valuations.}
It is well known that the greedy algorithm \GreedyGSVal, described in Algorithm~\ref{alg:gs-demand-query}, correctly solves a demand query for \GS\ valuations (see, e.g., \cite{gs-survey}).
The algorithm iteratively adds an element of a highest marginal utility, and terminates once all remaining elements have negative marginal utility.

Algorithm~\ref{alg:gs-demand-query-contract} builds on the observation of~\cite{DEFK21} that an agent's best response  to a contract $\alpha$ can be computed via a demand query. 
Observe that Algorithm~\ref{alg:gs-demand-query-contract} resolves ties in favor of outcomes with larger $f$ values,
unlike Algorithm~\ref{alg:gs-demand-query} which breaks ties arbitrarily.


\vspace{1em} 

\noindent
\begin{minipage}{0.48\linewidth}
    \begin{algorithm}[H]
\caption{$\GreedyGSVal(f, p)$}
\label{alg:gs-demand-query}
\begin{algorithmic}[1]
\State Initialize $S,S_0, S_1, \ldots, S_n \gets \emptyset$
\For{$i = 1,\ldots,n$}
    \State \label{st:alg-gs-demand-query-i} Let $x_i \in \argmax_{x\notin S}{ \big( f(x \mid S)-p(x) \big)}$
    \If {$ f(x_i \mid S)-p(x_i) \leq 0$}
        \State return S
    \EndIf
    \State $S_i \gets S_{i-1} + x_i$, $S \gets S_i$
\EndFor
\State return S
\Statex Break ties in step (\ref{st:alg-gs-demand-query-i}) consistently and \textbf{arbitrarily}.
\end{algorithmic}
\end{algorithm}
\end{minipage}
\hfill
\begin{minipage}{0.48\linewidth}
    \begin{algorithm}[H]
\caption{$\GreedyGSContract(\boldsymbol{\alpha}, f, p)$}
\label{alg:gs-demand-query-contract}
\begin{algorithmic}[1]
\State Initialize $\Sa^*,S_0, S_1, \ldots, S_n \gets \emptyset$
\For{$i = 1,\ldots,n$}
    \State \label{st:alg-gs-demand-query-contract-i} Let $x_i \in \argmax_{x\notin S}{ \big( \boldsymbol{\alpha} f(x \mid S)-p(x) \big)}$
    \If {$ \boldsymbol{\alpha} f(x_i \mid S)-p(x_i) \leq 0$}
        \State return $\Sa^*$
    \EndIf
    \State $S_i \gets S_{i-1} + x_i$, $\Sa^* \gets S_i$
\EndFor
\State return $\Sa^*$
\Statex Break ties in step (\ref{st:alg-gs-demand-query-contract-i}) consistently \textbf{in favor of the larger $f$ value.}

\end{algorithmic}
\end{algorithm}
\end{minipage}

\paragraph{\Ultra\ valuations.}
The demand oracle for \Ultra\ functions, established in~\cite{Leh17} and presented in Algorithm~\ref{alg:ultra-demand-query}, relies on the \wl\ property (Definition~\ref{def:well-layered}). The key difference between this algorithm and its counterpart for \GS\ functions (Algorithm~\ref{alg:gs-demand-query}) is that Algorithm~\ref{alg:ultra-demand-query} continues to add elements greedily even after encountering a negative marginal contribution. Indeed, the \wl\ property implies that this behavior is necessary to correctly compute a demand set for \Ultra\ functions. This is demonstrated in Example~\ref{ex:gs-vs-ultra}.

Using the connection between demand queries and the agent's best response once again, Algorithm~\ref{alg:ultra-demand-query-contract} computes the agent's best response for a contract $\alpha$, for \Ultra\ rewards, breaking ties in favor of higher $f$ values (see Lemma~\ref{lemma:contract-demand-query-ultra-add}).

\vspace{1em}
\noindent

\begin{minipage}{0.48\linewidth}
\begin{algorithm}[H]
\caption{$\GreedyUltraVal(f, p)$}
\label{alg:ultra-demand-query}

\begin{algorithmic}[1]
\State Initialize $S^*,S_0, S_1, \ldots, S_n \gets \emptyset$
\For{$i = 1,\ldots,n$}
    \State \label{st:alg-ultra-demand-query-i} 
     Let $\displaystyle x_i \in \argmax_{x\notin S_{i-1}}{ \big( f(x \mid S_{i-1})-p(x) \big)}$
    \State $S_i \gets S_{i-1} + x_i$
\EndFor
\State \label{st:alg-ultra-demand-query-final} Let $\displaystyle S^* \in \argmax_{(S_i)_{i \in [n]}}{\big( f(S_i)-p(S_i) \big)}$ 
\State \Return $S^*$
 \Statex Break ties in steps (\ref{st:alg-ultra-demand-query-i}) and (\ref{st:alg-ultra-demand-query-final}) consistently and \textbf{arbitrarily}.

\end{algorithmic}
\end{algorithm}
\end{minipage}
\hfill
\begin{minipage}{0.48\linewidth}
 \begin{algorithm}[H]
\caption{$\GreedyUltraContract(\boldsymbol{\alpha}, f, p)$}
\label{alg:ultra-demand-query-contract}
\begin{algorithmic}[1]
\State Initialize $\Sa^*,S_0, S_1, \ldots, S_n \gets \emptyset$
\For{$i = 1,\ldots,n$}
    \State \label{st:alg-ultra-demand-query-contract-i} 
    Let $\displaystyle  x_i \in \argmax_{x\notin S_{i-1}}{ \big(\boldsymbol{\alpha} f(x \mid S_{i-1})-p(x) \big)}$
    \State $S_i \gets S_{i-1} + x_i$
\EndFor
\State \label{st:alg-ultra-demand-query-contract-final} Let $\displaystyle \Sa^* \in \argmax_{(S_i)_{i \in [n]}}{\big(\boldsymbol{\alpha} f(S_i)-p(S_i) \big)}$ 
\State \Return $\Sa^*$
 \Statex Break ties in steps (\ref{st:alg-ultra-demand-query-contract-i}) and (\ref{st:alg-ultra-demand-query-contract-final}) consistently \textbf{in favor of the larger $f$ value.}
\end{algorithmic}
\end{algorithm}
\end{minipage}

\medskip

The following example demonstrates that it is necessary to continue adding elements even after all marginal utilities are negative in order to correctly answer a demand query for \Ultra\ functions.

\noindent 
This explains the transition from \GreedyGSVal\ to \GreedyUltraVal.

\begin{example}[\GS\ vs. \Ultra\ demand queries]
\label{ex:gs-vs-ultra}
Consider the set function $f$ over two elements, where $f(\{1\})=f(\{2\})=0$ and $f(\{1,2\})=10$.
One can easily verify that $f$ is \Ultra\ (e.g., by the Exchange property) but not \gs\ (e.g., by observing it is not submodular). 
Let the prices be $p_1=p_2=1$.
We first consider $\GreedyGSVal$. 
Since the marginal utility of each individual element is negative, the algorithm terminates immediately and returns the empty set.

In contrast, \GreedyUltraVal\ adds element 1 (without loss of generality) despite its negative marginal utility, yielding the set $S_1=\{1\}$ with utility $0-1=-1$. 
In the second iteration, it adds element 2, generating $S_2=\{1,2\}$  with utility $10-2=8$. 
The algorithm then returns $S_2$, which has the highest utility among $S_0,S_1$ and $S_2$, and is indeed the correct demand for $f$ and prices $p$.
\end{example}

\subsection{Preliminaries for Optimal Contract Computation}
\label{sec:optimal-contract-computation}

Although the optimal contract can be any real value in $[0,1]$, a key result from \cite{DEFK21} states that 
for any monotone reward function $f$ and cost function $c$,
there are only finitely many candidates for the optimal contract.
These are the values of $\alpha$ where the agent's best response shifts between action sets. 
These values are called {\em critical values}, and the set of all such values is denoted by $C_{f,c}$.
The following lemma establishes $f(S_\alpha)$ and $c(S_\alpha)$ are monotone with respect to $\alpha$.

\begin{lemma}[Monotonicity lemma,  \cite{DFG24combinatorial}]
\label{lemma:monotonicity-of-critical-values}
Let $S_\alpha, S_\beta \subseteq A$ be two distinct sets that maximize the agent's utility under contracts $\alpha, \beta$, for $0 \leq \alpha < \beta \leq 1$. Then, $f(S_\alpha) < f(S_\beta)$ and $c(S_\alpha) < c(S_\beta)$.
\end{lemma}

The following theorem from \cite{DFG24combinatorial} establishes a sufficient condition for constructing a poly-time algorithm for the optimal contract problem.

\begin{theorem}
\cite{DFG24combinatorial}
\label{thm:optimal-contract}
    For any monotone reward function $f$ and cost function $c$, the following two conditions are jointly sufficient to solve the optimal contract problem in polynomial time:
    \begin{enumerate}[leftmargin=15pt,itemsep=1pt] 
        \item There is a $poly(n)$-time algorithm for computing the agent's best response for a given contract $\alpha$.
        \item The number of critical values is polynomial in $n$.
    \end{enumerate}
\end{theorem}

\section{Optimal Contract with Ultra Rewards (and Additive Costs)}
\label{sec:main-result} \label{sec:ultra-add}

In this section we present our main result --- a poly-time algorithm for the optimal contract problem under \Ultra\ rewards (and additive costs).

\begin{theorem}
\label{thm:optimal-contract-ultra-add}
Let $f:\setfunc$ be an \Ultra\ reward function and $c:\setfunc$ an additive cost function. 
Then the optimal contract $\alpha$ can be computed in polynomial time.
\end{theorem}

To prove Theorem~\ref{thm:optimal-contract-ultra-add}, we present an efficient algorithm for computing the agent's best response (Section~\ref{sec:ultra-add-demand-query}), and establish a polynomial upper bound on the size of the set of the critical values (Section~\ref{sec:ultra-add-critical-values}).
By Theorem~\ref{thm:optimal-contract}, these are sufficient conditions for solving the optimal contract problem efficiently. 

\subsection{Agent's Best Response}
\label{sec:ultra-add-demand-query}

The following lemma (whose proof is deferred to  Appendix~\ref{sec:appendix-ultra-add}) shows that Algorithm~\ref{alg:ultra-demand-query-contract} computes the agent's best response for \Ultra\ rewards and additive costs.


\begin{restatable}{lemma}{contractdemandqueryultraadd}
\label{lemma:contract-demand-query-ultra-add}
Let $f:\setfunc$ be an \Ultra\ reward function, $p:\setfunc$ an additive cost function, and $\alpha\in[0,1]$. 
Then Algorithm~\ref{alg:ultra-demand-query-contract} computes the agent's best response; that is, \\
$
    \GreedyUltraContract(\alpha, f, p) \in D^*_{f,p}(\alpha)
$.
\end{restatable}


\subsection{Poly-Many Critical Values}
\label{sec:ultra-add-critical-values}

The following theorem is the key result in this paper.

\begin{theorem} \label{thm:ultra-add-critical-values} Let $f$ be a monotone \Ultra\ reward function and let $c:\setfunc$ be an additive cost function. It holds that  $ |C_{f,c}| \leq \frac{n(n+1)}{2} $.
\end{theorem}

Before presenting the formal proof of Theorem~\ref{thm:ultra-add-critical-values}, we outline its main ideas.
We begin by defining a notion of a {\em generic} additive cost function with respect to \Ultra\ rewards, designed so that Algorithm~\ref{alg:ultra-demand-query-contract} (which computes the agent's best response) encounters at most one tie-breaking event.
We then show that under generic costs, as the contract $\alpha$ increases, the agent's best response changes in one of two ways: either a set of actions is added to the best response, or a cheaper action is replaced by a more expensive one. Using a potential function argument, we prove that the number of such changes is polynomially bounded. This yields a polynomial bound on the number of critical values.
Finally, we extend the result from generic to arbitrary additive costs.

\subsubsection{Generic Cost Function}

We begin by defining the notion of genericity. A cost function is {\em generic} with respect to an \Ultra\ reward function, if computing the agent's best response involves at most one tie-breaking event.

\begin{definition}
 \label{def:generic_cost_function}
Given a function $f$, an additive cost function $c:\setfunc$, and an unordered pair of disjoint sets\footnote{At most one of $T_1,T_2$ can be $\emptyset$, where $f(\emptyset \mid S) = 0$ and $c(\emptyset)=0$ for any $S$. } 
$T_1,T_2\subseteq A$ 
, we define
\[
\Gamma_{f,c}(T_1,T_2) = \{\alpha  \mid  \exists S_1,S_2\subseteq  A \mbox{ s.t. } \alpha f(T_1 \mid S_1)-c(T_1)=\alpha f(T_2 \mid S_2)-c(T_2)\}
\]
\end{definition}
That is, $\Gamma_{f,c}(T_1, T_2)$ is the set of all values of $\alpha$ for which the marginal utility of $T_1$ with respect to some set equals the marginal utility of $T_2$ with respect to another (possibly different) set.
Any pair $T_1, T_2$ such that $\alpha \in \Gamma_{f,c}(T_1, T_2)$ represents a potential tie-breaking case in the execution of \GreedyUltraContract. Such ties may arise either during an iteration, when the agent is indifferent between adding one action or another, or in the final step, due to indifference between two sets $S_i$ and $S_j$.

\begin{definition}
 \label{def:generic_cost} (Generic cost)
An additive cost function $c$ is said to be {\em generic} with respect to a reward function $f$ if for every $\alpha > 0$, there exists at most one (unordered) pair of disjoint action sets $T_1,T_2 \subseteq A$ such that $\alpha \in \Gamma_{f,c}(T_1, T_2)$.
\end{definition}

Notably, for generic costs, the cost of any set of actions is unique. The next observation (whose proof is deferred to  Appendix~\ref{sec:appendix-ultra-add}) states this formally.

\begin{restatable}{observation}{genericcostdistinctcosts}
\label{obs:generic-cost-distinct-costs}
Let $c$ be a cost function that is generic with respect to a reward function $f$. 
Then every action set has a unique cost. In particular, the cost of any action set is nonzero, and the cost of each individual action is unique.
\end{restatable}


The following proposition shows that under generic costs, the size of the critical values set is polynomially bounded.

\begin{proposition}
\label{prop:generic-cost-ultra-critical-values}
Let $f:\setfunc$ be an \Ultra\ reward function and $c:\setfunc$ an additive set function that is {\em generic} with respect to $f$. Then 
$|C_{f,c}| \le n(n+1)/2$.
\end{proposition}

To prove Proposition~\ref{prop:generic-cost-ultra-critical-values}, we follow a potential function argument, as in \cite{DEFK21}. Specifically,
we define a {\em potential} function $\Phi:2^A \to \mathbb{Z}_{\geq 0}$ and show that $\Phi$ is bounded and that $\Phi(S_\alpha)$ increases with $\alpha$. 
We assign to each action $a \in A$ a rank $r_a$, where the most expensive action has rank $n$, the next has $n-1$, and so on. From Observation~\ref{obs:generic-cost-distinct-costs}, the cost of each action is unique, ensuring uniqueness of the rank. The potential of a set $S$ is then defined as $\Phi(S) = \sum_{a \in S} r_a$.
The proof of Proposition~\ref{prop:generic-cost-ultra-critical-values} now follows from the next lemma.

\begin{lemma}
 \label{lemma:potential_of_best_responses}
For every $\alpha', \alpha \in C_{f,c}$, $\alpha' < \alpha$, we have $\Phi(S_{\alpha'}) \leq \Phi(S_{\alpha}) - 1$.
\end{lemma}

Before proving the lemma, we show how it implies the bound on the size of the critical values set (Proposition~\ref{prop:generic-cost-ultra-critical-values}).
Let $C_{f,c}=\{\alpha_1, \ldots, \alpha_k\}$, where $\alpha_1 < \cdots < \alpha_k$.
Note that $\Phi(S_{\alpha_1}) \geq 1$ since $S_{\alpha_1} \neq \emptyset$.
By the fact that $\Phi(S_{\alpha_j+1}) \geq \Phi(S_{\alpha_j}) + 1$ for all $j$, it follows that $\Phi(S_{\alpha_k}) \geq k$. 
On the other hand, the potential is bounded above by the sum of all ranks:
$\Phi(S_{\alpha_k}) \leq \sum_{a \in A} r_a = \sum_{i=1}^{n} i = n(n+1)/2$. 
Putting all together, we conclude that $ \lvert C_{f,c} \rvert = k \leq n(n+1)/2$.

To prove Lemma~\ref{lemma:potential_of_best_responses}, it suffices to prove that $\Phi(S_{\alpha'}) < \Phi(S_{\alpha})$ for all neighboring $\alpha', \alpha \in C_{f,c}$, where $\alpha' < \alpha$ are neighbors if $(\alpha', \alpha) \cap C_{f,c} = \emptyset$. 
To this end, Proposition~\ref{prop:neighboring_critical_values} shows that as $\alpha$ increases,
the best response $S_\alpha$ either gains a set of actions or replaces a single cheaper action with a more expensive one.
In both cases, we have $\Phi(S_{\alpha'}) \leq \Phi(S_{\alpha}) - 1$, as desired.

\begin{proposition}
\label{prop:neighboring_critical_values}
Let $f$ be an \Ultra\ reward function and $c$ a generic additive cost function. 
For any neighboring critical values $\alpha' < \alpha$, the set $S_{\alpha'}$ takes one of the following two forms:
\begin{enumerate}
    \item $S_{\alpha'} = S_{\alpha} \setminus T$ for some $T \subseteq S_{\alpha}$, or
    \item $S_{\alpha'} = (S_{\alpha} \setminus \{a_1\}) \cup \{a_2\}$ for some $a_1 \in S_{\alpha}$ and $a_2 \notin S_{\alpha}$, where $c(a_2) < c(a_1)$.
\end{enumerate}
\end{proposition}

The proof of Proposition~\ref{prop:neighboring_critical_values} is the core of this section and relies on the next lemma.

\begin{lemma}
\label{lemma:critical-implies-tie-breaking}
Let $f:\setfunc$ be an \Ultra\ reward function, $c:\setfunc$ an additive cost function that is generic with respect to $f$, and $\alpha \in C_{f,c}$ a critical value. Then the computation of $\GreedyUltraContract(\alpha, f, c)$ encounters a tie-breaking. Moreover, one of the following holds:
\begin{enumerate}
    \item $\alpha \in \Gamma_{f,c}(S_i \setminus S_j, \emptyset)$ for some $j < i$, or
    \item $\alpha \in \Gamma_{f,c}(\{a\}, \{a'\})$ for some distinct actions $a,a' \in A$.
\end{enumerate}
\end{lemma}

\begin{proof}
We begin the proof by showing that if $\alpha$ is critical, then $\GreedyUltraContract(\alpha, f, c)$ must encounter a tie-breaking. Assume towards contradiction that there are no tie-breaks. 
Thus, there is no tie-breaking in step (\ref{st:alg-ultra-demand-query-contract-i}) when choosing $x_i$ for any $i$. That is, for every $i$ and for every $x\notin S_{i-1}$ and $x\neq x_i$ it holds that
\[
\alpha f(x \mid S_{i-1})-c(x)<
\alpha f(x_i \mid S_{i-1}) - c(x_i).
\]
Similarly, there is no tie-breaking when choosing $\Sa^*$ in step
(\ref{st:alg-ultra-demand-query-contract-final}). 
So for every $i$ s.t. $S_i \neq S_\alpha ^* $ we have
\[
\alpha f(S_i) -c(S_i) < 
\alpha f(S_\alpha ^*) - c(S_\alpha ^*).
\]
Since these are finitely many strict inequalities, there exists a sufficiently small $\epsilon>0$ such that, for every $i$ and every $x \notin S_{i-1}$ with $x \neq x_i$, we also have
\[
(\alpha-\epsilon) f(x \mid S_{i-1})-c(x) <
(\alpha-\epsilon) f(x_i \mid S_{i-1}) - c(x_i),
\] 
and for every $i$ s.t. $S_i \neq \Sa^*$ we also have
\[
(\alpha-\epsilon) f(S_i) -c(S_i) < 
(\alpha-\epsilon) f(S_\alpha ^*) - c(S_\alpha ^*).
\]
Consequently, executing \GreedyUltraContract\ on contracts $\alpha$ and $(\alpha-\epsilon)$ would follow the same path and return the same result, contradicting the assumption that $\alpha$ is a critical value.
Therefore, a tie-breaking must occur, either in step~(\ref{st:alg-ultra-demand-query-contract-i}), when selecting $x_i$ for some $i$, or in the final step~(\ref{st:alg-ultra-demand-query-contract-final}), when selecting $\Sa^*$.
A tie-breaking in step~(\ref{st:alg-ultra-demand-query-contract-i}) for some $i$ occurs if and only if there exists some $x \notin S_{i-1}$ with $x \neq x_i$ such that
$\alpha f(x \mid S_{i-1}) - c(x) = \alpha f(x_i \mid S_{i-1}) - c(x_i)$
implying that $\alpha \in \Gamma_{f,c}({x_i}, {x})$.
A tie-breaking in the final step~(\ref{st:alg-ultra-demand-query-contract-final}) occurs if and only if there exist some $j < i$ such that
$\alpha f(S_i)-c(S_i) = \alpha f(S_j)-c(S_j)$.
Subtracting $\alpha f(S_j)-c(S_j)$ from both sides yields $\alpha f((S_i \setminus S_j) \mid S_j)-c(S_i \setminus S_j) = 0$, 
and hence $\alpha \in \Gamma_{f,c}(S_i\setminus S_j, \emptyset)$.
\end{proof}

We are now ready to prove Proposition~\ref{prop:neighboring_critical_values}. 
Notably, the proof approach in \cite{DEFK21} does not carry over as is, due to the differences between \GreedyGSContract\ and \GreedyUltraContract.

\begin{proof}[Proof of Proposition~\ref{prop:neighboring_critical_values}]
Let $f$ be an \Ultra\ reward function, $c$ be an additive cost function that is generic with respect to $f$, and $\alpha\in C_{f,c}$.
Let $x_1^\alpha, \ldots, x_n^\alpha$ and $S_1^\alpha, \ldots, S_n^\alpha$ denote the sequence of choices produced by $\GreedyUltraContract(\alpha,f,c)$, and let  $\Sa^*$ denote its result. 
Similarly, let $x_1^{\alpha-\epsilon}, \ldots, x_n^{\alpha-\epsilon}$ and $S_1^{\alpha-\epsilon}, \ldots, S_n^{\alpha-\epsilon}$ denote the sequence produced by $\GreedyUltraContract((\alpha-\epsilon),f,c)$ for some $\epsilon$, and let $\Sae^*$ denote its result. 
Once we establish that $x_i^\alpha=x_i^{\alpha-\epsilon}$ for some 
$i$, we denote this value by $x_i$ from that point on. Similarly, if 
$S_i^\alpha=S_i^{\alpha-\epsilon}$ for some $i$, we denote it by 
$S_i$ thereafter.

Without loss of generality, let $A=\{a_1, \ldots, a_n\}$ be ordered such that $x^\alpha_i=a_i$ for every $i$; that is, when running $\GreedyUltraContract(\alpha,f,c)$, $a_i$ is selected in iteration $i$.
By Lemma~\ref{lemma:critical-implies-tie-breaking}, there must be a tie-breaking in the computation of $\GreedyUltraContract(\alpha,f,c)$. Also by this lemma, 
either $\alpha\in\Gamma_{f,c}((S^\alpha_i\setminus S^\alpha_j), \emptyset)$ for some $j<i$ or $\alpha\in\Gamma_{f,c}(\{a_k\},\{a_\ell\})$ for some $k<\ell $. We distinguish between these two cases. 

\textbf{Case 1:} $\alpha \in \Gamma_{f,c}((S^\alpha_i\setminus S^\alpha_j), \emptyset)$ for some $j<i$.
By genericity, for every $k$ and every $x\notin S^\alpha_{k-1}$ with $x\neq a_k$, it holds that $\alpha \notin \Gamma_{f,c}(\{a_k\}, \{x\})$. Therefore, there is no tie-breaking in step (\ref{st:alg-ultra-demand-query-contract-i}) of Algorithm~\ref{alg:ultra-demand-query-contract} in any iteration $i$. Thus,
\begin{equation}
\label{eq:case_1_no_tie_loop}
\alpha f(x \mid S^\alpha_{k-1}) - c(x) <
\alpha f(a_k \mid S^\alpha_{k-1}) - c(a_k)    
\end{equation}

Also by genericity, for any $i'<j'$ (that are distinct from $i,j$), $\alpha \notin \Gamma_{f,c}((S^\alpha_{i'} \setminus S^\alpha_{j'}), \emptyset)$, so $\alpha f(S^\alpha_{i'})-c(S^\alpha_{i'}) \neq \alpha f(S^\alpha_{j'})-c(S^\alpha_{j'})$. Therefore, the tie-breaking could only have been between $S^\alpha_i$ and $S^\alpha_j$, so they must have an equal utility, namely $\alpha f(S^\alpha_i)-c(S^\alpha_i) = \alpha f(S^\alpha_j)-c(S^\alpha_j)$.

Moreover, we show that since $\alpha$ is critical, it must be that $S_\alpha^* \in \{S^\alpha_i,S^\alpha_j\}$. To this end, assume that it is not the case, thus for every $k$ such that $S^\alpha_k \neq S_\alpha^*$ it holds that
\begin{equation}
\label{eq:case_1_no_tie_final}
\alpha f(S^\alpha_k)-c(S^\alpha_k) < 
\alpha f(S_\alpha^*)-c(S_\alpha^*)
\end{equation}
Since there are finitely many strict inequalities of types (\ref{eq:case_1_no_tie_loop}), (\ref{eq:case_1_no_tie_final}), there exists a sufficiently small $\epsilon>0$ such that, for every $k$ and for every $x\notin S^\alpha_{k-1}$ with $x\neq a_k$, we also have
\[
(\alpha-\epsilon) f(x \mid S^\alpha_{k-1}) - c(x) <
(\alpha-\epsilon) f(a_k \mid S^\alpha_{k-1}) - c(a_k),
\]
and for every $k$ such that $S^\alpha_k\neq  S_\alpha^*$ we also have
\[
(\alpha-\epsilon) f(S^\alpha_k)-c(S^\alpha_k) < 
(\alpha-\epsilon) f(S_\alpha^*)-c(S_\alpha^*).
\]    
Consequently, executing \GreedyUltraContract\ on contracts $\alpha$ and $(\alpha-\epsilon)$ would follow the same path and return the same result, contradicting the assumption that $\alpha$ is a critical value.
As a result, it holds that $S_\alpha^* \in \{S^\alpha_i,S^\alpha_j\}$. 
Moreover, by the monotonicity of $f$, $f(S^\alpha_i) \geq f(S^\alpha_j)$, 
and since ties are broken in favor of the larger $f$ values, it follows that $S_\alpha^* = S^\alpha_i$.

Now, for every $k\notin \{i,j\}$ it holds that
\begin{equation}
\label{eq:case_1_no_tie_final_2}
    \alpha f(S_k)-c(S_k) < 
\alpha  f(S^\alpha_i)-c(S^\alpha_i) =
\alpha f(S^\alpha_j)-c(S^\alpha_j).
\end{equation}
Since there are finitely many inequalities of types (\ref{eq:case_1_no_tie_loop}), (\ref{eq:case_1_no_tie_final_2}), there exists a sufficiently small $\epsilon>0$ such that, for every $k$ and every $x\notin S^\alpha_{k-1}$ with $x\neq a_k$, we have
\[
(\alpha-\epsilon) f(x \mid S^\alpha_{k-1}) - c(x) <
(\alpha-\epsilon) f(a_k \mid S^\alpha_{k-1}) - c(a_k),
\]
and for every $k\notin{i,j}$ we also have
\begin{equation}
  \label{eq:case_1_i_j_are_best}  
(\alpha-\epsilon) f(S^\alpha_k)-c(S^\alpha_k) < 
(\alpha-\epsilon) f(S^\alpha_j)-c(S^\alpha_j).
\end{equation}
Moreover, from $f(S^{\alpha}_i) \geq f(S^{\alpha}_j)$ and $\alpha f(S^{\alpha}_i)-c(S^{\alpha}_i) = \alpha f(S^{\alpha}_j)-c(S^{\alpha}_j)$ we get $c(S^{\alpha}_i) \geq c(S^{\alpha}_j)$. But from Observation~\ref{obs:generic-cost-distinct-costs}, the cost of a set is unique, and therefore $c(S^{\alpha}_i) > c(S^{\alpha}_j)$ and $f(S^{\alpha}_i) > f(S^{\alpha}_j)$. We get: 
$
(\alpha-\epsilon) f(S^{\alpha}_i)-c(S^{\alpha}_i) <
(\alpha-\epsilon) f(S^{\alpha}_j)-c(S^{\alpha}_j).
$

Consequently, executing \GreedyUltraContract\ on contracts $\alpha$ and $(\alpha-\epsilon)$ would produce the same sequences $x_1^\alpha, \ldots, x_n^\alpha$ and $S_1^\alpha, \ldots, S_n^\alpha$, but then return $S_{\alpha-\epsilon}^* = S^{\alpha}_j$.
To conclude, the neighbor critical value $\alpha' < \alpha$ has $S_{\alpha'}^*=S_{\alpha-\epsilon}^*= S_\alpha^* - T$ for $T = (S^{\alpha}_i\setminus S^{\alpha}_j)$. 

\textbf{Case 2:} $\alpha \in \Gamma (\{a_k\},\{a_\ell\})$ for some $k<\ell$. Recall that by Lemma~\ref{lemma:critical-implies-tie-breaking}, there has to be a tie-breaking in the computation of $\GreedyUltraContract(\alpha, f,c)$. By genericity, for every $i<j$, it holds that $\alpha\notin\Gamma_{f,c}(S_i\setminus S_j, \emptyset)$. Thus, $\alpha f(S_i)-c(S_i) \neq \alpha f(S_j)-c(S_j)$.
Therefore, for every $i$ s.t. $S_\alpha^*\neq S_i$: 
\begin{equation}
\label{eq:no_tiebreaks_in_last} 
\alpha f(S_i) - c(S_i) < 
\alpha f(S_\alpha^*) - c(S_\alpha^*)
\end{equation}
It follows that there is no tie-breaking in the final step (\ref{st:alg-ultra-demand-query-contract-final}) when choosing $\Sa^*$. Therefore, the tie-breaking must occur in step (\ref{st:alg-ultra-demand-query-contract-i}) for some $i$. Moreover, by genericity, for every $i$ and every $x\notin S_{i-1}$ with $x\neq a_i$, and --- if $i=k$ --- also $x\neq a_\ell$, we have that $\alpha \notin \Gamma_{f,c}(\{x\},\{a_i\})$. This implies that
\begin{equation}
\label{eq:no_tiebreaks_in_i} 
\alpha f(x \mid S_{i-1}) - c(x) <
\alpha f(a_i \mid S_{i-1}) - c(a_i)
\end{equation}
Therefore, the tie-breaking could only be between $a_k$ and $a_\ell$ in iteration $k$, so they must have an equal marginal utility, i.e.  
$\alpha f(a_k \mid S_{k-1})-c(a_k) = 
 \alpha f(a_\ell \mid S_{k-1})-c(a_\ell)$. 
Since tie-breaking favors the larger $f$ value (equivalently, the larger cost), so $c(a_k) \geq c(a_\ell)$. 
But from Observation~\ref{obs:generic-cost-distinct-costs}, action costs are unique, therefore $c(a_k) > c(a_\ell)$ and $f(a_k \mid S_{k-1}) > f(a_\ell \mid S_{k-1})$.

Observe that for every $k<i<\ell$, $a_i$ was chosen, and specifically chosen over $a_\ell$ with no tie-breaking, so:
\[
\alpha f(a_\ell \mid S_{i-1}) - c(a_\ell) <
\alpha f(a_i \mid S_{i-1}) - c(a_i)
\]
By rearranging (adding $\alpha f(S_{i-1}) - c(S_{i-1})$ to both sides) we get the following
\begin{equation}
\label{eq:Si_is_better_than_Sl}
\alpha f(a_\ell + S_{i-1}) - c(a_\ell + S_{i-1}) <
\alpha f(a_i + S_{i-1}) - c(a_i + S_{i-1})
\end{equation}

In addition, following from the \wl\ property (Definition~\ref{def:well-layered}),  for every $i$, $S_i$ is best $i$-sized response for $\alpha$. 
%
Using this fact, we get that for every $k<i<\ell$ and every $r>i$ s.t. $r\neq \ell$, 
the following holds:
\begin{equation} \label{eq:Si_is_better_than_replacing_ak_al}
    \alpha f(S_{i-1}+a_i) -c(S_{i-1}+a_i) > 
    \alpha f((S_{i-1}-a_k+a_\ell)+a_r) - 
    c((S_{i-1}-a_k+a_\ell)+a_r)  
\end{equation}
Notably, $S_{i}=S_{i-1}+a_i$. Moreover, for any $k<i<\ell$, $a_k$ is a member of $S_{i-1}$ and $a_\ell$ is not, therefore $(S_{i-1}-a_k+a_\ell)+a_r$ is a set of size $i$. 
$S_i$ is best $i$-sized response to $\alpha$, so:
\[
\alpha f(S_{i-1}+a_i) -c(S_{i-1}+a_i) =
\alpha f(S_{i-1} -a_k +a_k +a_i) -c(S_{i-1} -a_k +a_k 
 +a_i)
\]\[
\geq 
\alpha f((S_{i-1}-a_k+a_\ell)+a_r)-c((S_{i-1}-a_k+a_\ell)+a_r)
\]
By rearranging (specifically, by subtracting $\alpha f(S_{i-1}-a_k)+c(S_{i-1}-a_k)$ from both sides), we get:
\[
\alpha f(a_k+a_i \mid S_{i-1}-a_k) -c(a_k+a_i) \geq \alpha f(a_\ell+a_r \mid S_{i-1}-a_k)-c(a_\ell+a_r) 
\]
We next show that the last inequality must be strict. Indeed, since $r>i$, it holds that $a_r\neq a_i$; therefore $\{a_k,a_i\}$ and $\{a_\ell,a_r\}$ are disjoint. Thus, by genericity, $\alpha \notin \Gamma (\{a_k,a_i\},\{a_\ell,a_r\})$, and the claim follows. This concludes the reasoning for Equation~ (\ref{eq:Si_is_better_than_replacing_ak_al}).

Furthermore, in general, for every $i$ such that $S_i \neq S_\alpha^*$, following from (\ref{eq:no_tiebreaks_in_last}), it also holds that 
\begin{equation} \label{eq:Si_is_better_than_S}
\alpha f(S)-c(S) \leq 
\alpha f(S_i)-c(S_i)  <
\alpha f(S_\alpha^*)-c(S_\alpha^*)
\end{equation}

Since there are finitely many inequalities of types
(\ref{eq:no_tiebreaks_in_last}),(\ref{eq:no_tiebreaks_in_i}),(\ref{eq:Si_is_better_than_Sl}),(\ref{eq:Si_is_better_than_replacing_ak_al}) and
(\ref{eq:Si_is_better_than_S}), there exists a sufficiently small $\epsilon>0$ for which every such inequality also holds strictly when $\alpha$ is replaced by $(\alpha-\epsilon)$.

We can now analyze the \GreedyUltraContract\ computation on $((\alpha-\epsilon), f,c)$. We first show that the sequences produced by this procedure would be
\begin{equation}
\begin{array}{cc}
     x_i^{\alpha-\epsilon}=
\begin{cases}
    a_i, & \text{if } i \neq k,\ell  \\
    a_\ell, & \text{if } i = k \\
    a_k, & \text{if } i = l
\end{cases}
     &  
S_i^{\alpha-\epsilon}=
\begin{cases}
    S_i^{\alpha}, & \text{if } i < k, i\geq\ell \\
    S_i^{\alpha} - a_k + a_\ell, & \text{if } k \leq i < \ell. 
\end{cases}
\end{array}
\label{eq:result-alpha-epsilon}
\tag{$\star$}
\end{equation}
We prove this by going over the iterations, for $i=1, \ldots, n$.

\begin{enumerate}[
  label=\arabic*.,
  wide=0pt,
  labelsep=0.5em,
  leftmargin=0pt,
  listparindent=-\labelwidth,
  itemsep=0pt,
  topsep=0pt
]
\item[$\textbf{For iterations }\boldsymbol{i<k}$:] For every $x\notin S_{i-1}$ with $x\neq a_i$, by (\ref{eq:no_tiebreaks_in_i}), we have $
(\alpha-\epsilon) f(x \mid S_{i-1}) - c(x) < 
(\alpha-\epsilon) f(a_i \mid S_{i-1}) - c(a_i)
$,
so $x_i^{\alpha-\epsilon}=a_i=x_i^{\alpha}$ and $S_{i}^{\alpha-\epsilon}=S_i^{\alpha}$, and we denote them by $x_i$ and $S_i$, respectively, hereafter.

\item[$\textbf{For iterations } \boldsymbol{i=k}$:] For every $x\notin S_{k-1}$ with $x\neq a_k$ and $x \neq a_\ell$, by (\ref{eq:no_tiebreaks_in_i}), we have 
\[
(\alpha-\epsilon) f(x \mid S_{k-1}) - c(x) <
(\alpha-\epsilon) f(a_k \mid S_{k-1}) - c(a_k) =
(\alpha-\epsilon) f(a_\ell \mid S_{k-1}) - c(a_\ell).
\]
And by $\alpha f(a_k \mid S_{k-1})-c(a_k) = \alpha f(a_\ell \mid S_{k-1})-c(a_\ell)$ and $f(a_k \mid S_{k-1}) > f(a_\ell \mid S_{k-1})$, 
\[
(\alpha-\epsilon) f(a_k \mid S_{k-1}) - c(a_k) <
(\alpha-\epsilon) f(a_\ell \mid S_{k-1}) - c(a_\ell).
\]
Therefore, $x_k^{\alpha-\epsilon}=a_\ell$ and $S_k^{\alpha-\epsilon}=S_{k-1} + a_\ell=S^\alpha_{k} - a_k + a_\ell$.

\item[$\textbf{For iterations } \boldsymbol{k<i<\ell}$ :] We show inductively that if $S^{\alpha-\epsilon}_{i'} = S^\alpha_{i'}-a_k+a_\ell$ for every $k\leq i'<i$, then $x_i^{\alpha-\epsilon}=a_i$ and $S^{\alpha-\epsilon}_{i} = S^\alpha_{i}-a_k+a_\ell$. (For the induction base, we have $S^{\alpha-\epsilon}_{k} = S^\alpha_{k}-a_k+a_\ell$).

By the induction hypothesis, for every $i'<i$ with $i'\neq k$, $a_{i'}\in S_{i-1}$ ($a_{i'}$ was already chosen in a previous iteration). Therefore, $x^{\alpha-\epsilon}$ can be one of $a_k$, $a_i$ or $a_r$ with $r>i$ and $r\neq \ell$. Let $S_i^{\alpha-\epsilon} = S_{i-1}^{\alpha-\epsilon} + a_r = (S_{i-1}^{\alpha} - a_k +a_\ell) + a_r$.

For $r>i$, ($r\neq \ell $), from (\ref{eq:Si_is_better_than_replacing_ak_al}) we have
\[
    (\alpha-\epsilon) f(S_{i-1} +a_i)-c(S_{i-1} +a_i) > 
    (\alpha-\epsilon) f((S_{i-1} -a_k +a_\ell) +a_r)-c((S_{i-1} - a_k + a_\ell) + a_r),
\]
and for $r=k$, from (\ref{eq:Si_is_better_than_Sl}), it holds that
\[
(\alpha-\epsilon) f(S_{i-1} +a_i)-c(S_{i-1} +a_i) > 
(\alpha-\epsilon) f(S_{i-1} +a_\ell)-c(S_{i-1} +a_\ell) = 
\]
\[
(\alpha-\epsilon) f((S_{i-1} -a_k +a_\ell) +a_k)-c((S_{i-1} - a_k + a_\ell) + a_k) 
\]
Therefore, $S_{i-1}^{\alpha-\epsilon}+a_r$ cannot be a best $i$-sized response to $(\alpha-\epsilon)$ when $r\neq i$. Consequently, $x_i^{\alpha-\epsilon} = a_i=x_i^\alpha$ and $S_i^{\alpha-\epsilon}=S_i^{\alpha}-a_k+a_\ell$.

\item[$\textbf{For iterations } \boldsymbol{i=\ell}$:] Let $x_\ell^{\alpha-\epsilon} = a_r$. It can either be that $r=k$, or $r>\ell$ (since $x_k^{\alpha-\epsilon}=a_\ell$ and $x_i^{\alpha-\epsilon}=a_i$ for $i<\ell$ and $i\neq k$). If $r>\ell $, then by (\ref{eq:Si_is_better_than_replacing_ak_al}), we have $
(\alpha-\epsilon) f(S_{\ell-1}^\alpha + a_\ell)-c(S_{\ell-1}^\alpha + a_\ell) >
(\alpha-\epsilon) f((S_{\ell-1}^\alpha - a_k +a_\ell) + a_r)-c((S_{\ell-1}^\alpha - a_k + a_\ell) + a_r).
$ 
Thus, $S_{\ell-1}^{\alpha-\epsilon}+a_r$ cannot be a best $\ell$-sized response for $r>\ell$, so it must be that $x_\ell^{\alpha-\epsilon}=a_k$, and $S^{\alpha-\epsilon}_\ell=S_{\ell-1}^{\alpha-\epsilon}+a_k = (S_{\ell}^\alpha - a_k + a_\ell)+a_k= S_{\ell}^\alpha$, and we denote it by $S_\ell$ hereafter.

\item[$\textbf{For iterations } \boldsymbol{i>\ell}$:] For every $x\notin S_{i-1}$ with $x\neq a_i$, by (\ref{eq:no_tiebreaks_in_i}), we have $(\alpha-\epsilon) f(x \mid S_{i-1}) -c(x) < 
(\alpha-\epsilon) f(a_i \mid S_{i-1}) -c(a_i)$, so $x_i^{\alpha-\epsilon}=a_i=x_i^\alpha$ and $S_i^{\alpha-\epsilon}=S_i^\alpha$, denoted by $x_i$ and $S_i$ hereafter.
\end{enumerate}

This concludes the proof that $\GreedyUltraContract((\alpha-\epsilon), f, c)$ yields the sequences given in Eq.~(\ref{eq:result-alpha-epsilon}).

It remains to show that $S_{\alpha-\epsilon}^* = S_\alpha^* - a_k +a_\ell$.
To this end, let $j$ be s.t. $S_\alpha^*=S^\alpha_j$.
From (\ref{eq:Si_is_better_than_S}), for every $i\neq j$, for every set $S$ of size $i$, it holds that $(\alpha-\epsilon) f(S) - c(S) < 
 (\alpha-\epsilon) f(\Sa^*) - c(\Sa^*)$. 
So in particular, for $S=S^{\alpha-\epsilon}_i$ (a set of size $i$), $
(\alpha-\epsilon) f(S^{\alpha-\epsilon}_i) - c(S^{\alpha-\epsilon}_i) < 
(\alpha-\epsilon) f(S^*_\alpha) - c(S^*_\alpha).
$
Consequently, $S^{\alpha-\epsilon}_i$ cannot be a best response for $(\alpha-\epsilon)$ for any $i\neq j$. Therefore, the only candidate left is $\Sae^*=S^{\alpha-\epsilon}_j$.
Recall that if $j<k$ or $j \geq \ell$, then $S_j^{\alpha-\epsilon}=S_j^\alpha$. So in this case, $\Sa^*=\Sae^*$, a contradiction to $\alpha$ being a critical value. 
Thus, $k \leq j < \ell$ must hold with $S_j^{\alpha-\epsilon}=S_j^\alpha-a_k+a_\ell$. As a result, $\Sa^*=\Sae^*-a_k+a_\ell$, and the neighbor critical value $\alpha' < \alpha$ has $S_{\alpha'}^*=S_{\alpha-\epsilon}^*= S_\alpha^* - a_k+a_\ell$, as desired. 
\end{proof}

\subsubsection{Arbitrary Cost Functions}

Next, we extend the bound on the size of the critical values set from generic to arbitrary cost functions. We show that a slight perturbation of any cost function results in a generic cost with probability 1. Together with a lemma from \cite{DEFK21}, which ensures that a sufficiently small perturbation does not decrease the number of the critical values, this implies the bound also holds for arbitrary additive costs.

\begin{definition} \label{def:epsilon-perturbation-of-c}
Given a cost function $c$, a cost function $\hat{c}$ is said to be an $\epsilon$-perturbation of $c$, if for
every action $a \in A$, $\hat{c}(a) \in [c(a), c(a) + \epsilon]$.
\end{definition}

\begin{lemma}[\cite{DEFK21}, Lemma 4.13] \label{lemma:epsilon-perturbation-critical-values}
For every monotone reward function $f$ and an additive cost function $c$, there exists an $\epsilon>0$
such that every $\epsilon$-perturbation $\hat{c}$ of $c$ has $ |C_{f,c}| \leq |C_{f,\hat{c}}| $.
\end{lemma}

Using this lemma we prove Theorem~\ref{thm:ultra-add-critical-values}, namely, an upper bound of $n(n+1)/2$ on the number of critical values for any \Ultra\ reward and additive cost functions.

\begin{proof}[Proof of Theorem~\ref{thm:ultra-add-critical-values}]
By Lemma~\ref{lemma:epsilon-perturbation-critical-values}, there exists $\epsilon>0$ such that every $\epsilon$-perturbation $\hat{c}$ of $c$ has $|C_{f,c}| \leq |C_{f,\hat{c}}|$. 
We show that when drawing $\hat{c}(a)$ uniformly at random from $[c(a), c(a) + \epsilon]$ for every action $a \in A$, the additive function $\hat{c}:\setfunc$ defined by $\hat{c}(S) = \sum_{a\in S} \hat{c}(a)$ will be generic with probability 1. 
As a result of Proposition~\ref{prop:generic-cost-ultra-critical-values}, $|C_{f,\hat{c}}| \leq n(n+1)/2$ for any generic $\hat{c}$.
Observe that the drawn $\hat{c}$ is not generic if the following event occurs with probability strictly greater than zero:
\[
\text{There exist two different (unordered) pairs } T_1,T_2 \text{ and } T_3,T_4 \text{ s.t. }
\Gamma_{f,\hat{c}}(T_1, T_2) \cap \Gamma_{f,\hat{c}}(T_3, T_4) \neq \emptyset\]
Indeed, in this case, there exists some $\alpha$ s.t. $\alpha \in \Gamma_{f,\hat{c}}(T_1, T_2)$ and $\alpha \in \Gamma_{f,\hat{c}}(T_3, T_4)$, a contradicting genericity.
We next show that this probability equals $0$. Applying the union bound, we get 
%
{\small
\[
\Pr\Big[ 
    \Gamma_{f,\hat{c}}(T_1, T_2) 
    \cap
    \Gamma_{f,\hat{c}}(T_3, T_4) \neq \emptyset \Big] \leq
\]
\[
\sum_{S_1,\, S_2,\, S_3,\, S_4} \Pr\Big[ 
    \exists\, \alpha > 0 \text{ : } 
    \alpha f(T_1 \mid S_1) - \hat{c}(T_1) 
      = 
    \alpha f(T_2 \mid S_2) - \hat{c}(T_2) 
    \text{  and  }
    \alpha f(T_3 \mid S_3) - \hat{c}(T_3) 
      = 
    \alpha f(T_4 \mid S_4) - \hat{c}(T_4)\Big]
\]}

If $f(T_1 \mid S_1) = f(T_2 \mid S_2)$, then 
$\Pr[\Gamma_{f,\hat{c}}(T_1, T_2) \cap \Gamma_{f,\hat{c}}(T_3, T_4) \neq \emptyset] \leq \Pr[\hat{c}(T_1) = \hat{c}(T_2)] = 0$.
The last transition is since $\hat{c}(T_1) = \sum_{a\in T_1} \hat{c}(a)$ and $\hat{c}(T_2) = \sum_{a\in T_2} \hat{c}(a)$ are the sums of two finite sets of i.i.d samples from the uniform distribution. Therefore, this event has measure zero. 

Otherwise, 
%
{\small\[
\Pr\Big[ 
\Gamma_{f,\hat{c}}(T_1, T_2) \cap \Gamma_{f,\hat{c}}(T3, T_4) \neq \emptyset\Big] 
    \leq 
\]
\[
    \sum_{S_1, S_2, S_3, S_4}\Pr\Big[ 
    \Big( f(T_1 \mid S_1) - f(T_2 \mid S_2) \Big) 
    \Big(\hat{c}(T_3) - \hat{c}(T_4) \Big) 
     = 
 \Big( f(T_3 \mid S_3) - f(T_4 \mid S_4)\Big) \Big(\hat{c}(T_1) - \hat{c}(T_2) \Big) 
    \Big] = 0,
\]}
where the last equality again holds since it's a measure-zero event.
Applying the union bound over these finitely many events yields that $\hat{c}$ is generic with probability 1. Altogether, we get
$|C_{f,c}| \leq
    |C_{f,\hat{c}}| \leq 
    \frac{n(n+1)}{2}$.
\end{proof}

\section{Beyond Additive Costs: Symmetric Plus Additive (\SPA)} \label{sec:beyond-additive}

In this section we extend our main result by showing that the optimal contract problem can be solved in polynomial time also for costs that are {\em a sum of an additive function and a monotone symmetric} function.

\begin{definition}[\SPA\ costs]
    A set function $c:\setfunc$ is called {\em Symmetric Plus Additive} (\SPA) if there exist a monotone symmetric set function $g:\setfunc$ (i.e. $g(S)=g(T)$ for every $S,T$ with $|S|=|T|$) and an additive set function $p:\setfunc$, such that for every set $S$, $c(S)=g(S)+p(S)$. 
\end{definition} 

Note that a \SPA\ set function is, in general, neither additive nor symmetric.

In the remainder of this section, whenever we write $c=g+p$, we mean that $c:\setfunc$ is the sum of an additive function $p:\setfunc$ and a monotone symmetric function $g:\setfunc$.

The following is the main result of this section.

\begin{theorem} \label{thm:ultra-spa}
    Given an \Ultra\ reward function and a \SPA\ cost function, the optimal contract can be computed in polynomial time.     
\end{theorem}

We prove Theorem~\ref{thm:ultra-spa} separately for \GS\ and \Ultra\ reward functions.
While the proof for \Ultra\ rewards also applies to \GS\ rewards as a special case, the \GS\ case admits a simpler proof worth presenting on its own.
In Section~\ref{sec:ft} we present a framework useful for exploring instances with \SPA\ cost function, and in Section~\ref{sec:ultra-spa}, we present the proof for \Ultra\ rewards. The proof for \GS\ rewards appears in Appendix~\ref{sec:appendix-gs-beyond-additive}.

\subsection{Best Response 
of Restricted Size and Truncated Set Functions} 
\label{sec:ft}

In this section we present a tool for bounding the number of critical values in instances with \GS\ or \Ultra\ rewards and \SPA\ costs.

We begin with an algorithm that computes the agent's best response for a monotone reward function $f$ (not necessarily \Ultra) and a \SPA\ cost function $c = g + p$, given as Algorithm~\ref{alg:demand-spa}.
In every iteration $i$ (see step~\ref{st:alg-demand-spa-i}), the algorithm selects a set that is a best response of size at most $i$.

\begin{minipage}{0.8\linewidth}
\begin{algorithm}[H]
\caption{$\DemandForSPA(\alpha, f, p, g)$}
\label{alg:demand-spa}
\begin{algorithmic}[1]
\State Initialize $S_0, S_1, \ldots, S_n \gets \emptyset$
\For{$i = 1,\ldots,n$}
    \State \label{st:alg-demand-spa-i} Let $S_i \in D^{\leq i}_{f,p}(\alpha) = \argmax_{S,|S| \leq i}\big( \alpha f(S)-p(S) \big)$
\EndFor
\State \label{st:alg-demand-spa-final} Let $S_\alpha ^* \in \argmax_{(S_i)_{i \in [n]}}{\big( \alpha f(S_i)-p(S_i)-g(S_i) \big)}$ 
\State \label{st:alg-demand-spa-last}\label{st:alg-demand-spa-return} \Return $S_\alpha^*$
\Statex Break ties in steps (\ref{st:alg-demand-spa-i}) and (\ref{st:alg-demand-spa-final}) consistently in favor of the larger $f$ value. 
\end{algorithmic}
\end{algorithm}
\end{minipage}




\medskip
The following lemma establishes the correctness of Algorithm~\ref{alg:demand-spa}. Its proof appears in Appendix~\ref{sec:appendix-spa-ft}.

\begin{restatable}
{lemma}
{demandforspa}
\label{lemma:demand-for-spa}

Let $f:\setfunc$ be a monotone set function, $c=g+p$ a \SPA\ cost function, and $\alpha \in [0,1]$ a contract. Assume that for every $i\in [n]$, there exists some $\epsilon_i>0$, such that for every $0<\epsilon\leq \epsilon_i$, step (\ref{st:alg-demand-spa-i}) of Algorithm~\ref{alg:demand-spa} generates the same result for $\alpha$ and $(\alpha+\epsilon)$. Then, 
$\DemandForSPA(\alpha, f, p, g)\in D^*_{f,c}(\alpha)$.

\end{restatable}


\begin{proof}
Let $S_1, \ldots, S_n$ denote the sequence produced by $\DemandForSPA(\alpha, f, p, g)$, and let $\Sa^*$ denote its result.
To prove the lemma, we need to show that the following two conditions hold: $(i)$ $\Sa^* \in D_{f,p}(\alpha)$, and $(ii)$ $f(\Sa^*) \geq f(S')$ for every $S'\in D_{f,p}(\alpha)$.

To show $(i)$ (i.e., $\Sa^* \in D_{f,c}(\alpha)$), we prove that $\Sa^*$ attains the highest utility among all sets. 
Let $S$ be any set with $k=|S|$ and let $k'=|S_k|$. 
Let $g_i$ be the value of $g$ on a set of size $i$.
By assumption, $S_k \in D^{\leq k}_{f,p}(\alpha)$, and since $|S_k|=k'\leq k$, we have that $g_{k'} \leq g_k$ by monotonicity of $g$. 
Thus,
\[
    f(S)-p(S)-g(S) = 
    f(S)-p(S)-g_k \leq 
    f(S_k)-p(S_k)-g_k \leq
\]\[
    f(S_k)-p(S_k)-g_{k'} =
    f(S_k)-p(S_k)-g(S_k) \leq 
    f(\Sa^*)-p(\Sa^*)-g(\Sa^*).    
\]
The proof of condition $(ii)$ appears in Lemma~\ref{lemma:demand-spa-maximal-f} in Appendix~\ref{sec:appendix-spa-ft}.
\end{proof}

We next introduce a family of set functions, called {\em truncated set functions}, which will be used to compute a best response up to a given size.

\begin{definition}[truncated function] \label{def:f_t}
Given a monotone function $f:\setfunc$, for every $1\leq t \leq n$, we define the {\em $t$-truncation} of $f$, $f_t:\setfunc$, to be:
\vspace{-4pt} 
\[
    f_t(S) = \begin{cases}
        f(S), & |S| \leq t \\
        \max \{f(S') \mid  S' \subset S, |S'|=t\}, & |S| > t
    \end{cases}
\]
\end{definition}

The following observation provides structural insights about truncated functions. The proof is presented in Appendix~\ref{sec:appendix-spa-ft}.

\begin{restatable}{observation}{ftstructuralinsights} \label{obs:f_t-structural-insights}
For every monotone set function $f:\setfunc$, $t\in [n]$, and an additive function $p:\setfunc$, it holds that $f_t$ is monotone, $f_t(S) \leq f(S)$ for every $S\subA$, and:
    \begin{enumerate}
    [ leftmargin=15pt,
    itemsep=0pt, 
    topsep=0pt]
    \item \label{en:f_t-structural-insights-1} If $S^* \in \argmax_{S,|S|\leq t} \Big( f(S)-p(S)\Big)$, then $S^* \in D_{f_t}(p)$. 
    \item \label{en:f_t-structural-insights-2} In particular, if $S^*\in D_f(p)$ and $|S^*|\leq t$, then $S^*\in D_{f_t}(p)$ as well.
    \item \label{en:f_t-structural-insights-3} If $S_t\in \argmax_{S,|S|=t} \Big( f(S)-p(S)\Big)$, then $S_t\in \argmax_{S, |S|\geq t} \Big( f_t(S)-p(S)\Big)$.
    \item \label{en:f_t-structural-insights-4} If $S^* \in D_{f_t}(p)$ and $|S^*|\leq t$, then $S^* \in \argmax_{S,|S|\leq t} \Big( f(S)-p(S)\Big)$.
\end{enumerate}
\end{restatable}
\vspace{-4pt}

An immediate corollary of Observation~\ref{obs:f_t-structural-insights} (\ref{en:f_t-structural-insights-4}) is that for any $t\in [n]$, every set in the demand of $f_t$ that is of size at most $t$ is also a best response of size at most $t$ with respect to $f$. 

\begin{corollary} \label{cor:f_t-demand-up-to-t}
If $\Sa^* \in D_{f_t,p}(\alpha)$ and $|\Sa^*| \leq t$, then $\Sa^* \in D^{\leq t}_{f,p}(\alpha)$. 
\end{corollary}

This corollary implies that to implement iteration $i$ of Algorithm~\ref{alg:demand-spa} (step~\ref{st:alg-demand-spa-i}), it suffices to find a set of size at most $i$ in the demand of $f_i$, for every $i\in [n]$.
This enables computing the optimal contract for \GS\ and \Ultra\ rewards under \SPA\ costs.

\subsection{\Ultra\ Rewards and \SPA\ Costs} \label{sec:ultra-spa} 

This section proves Theorem~\ref{thm:ultra-spa}. 
Section~\ref{sec:ultra-spa-demand} presents a poly-time algorithm for computing an agent's best response  
and Section~\ref{sec:ultra-spa-critical-values} shows that the number of critical values is polynomially bounded.
These two conditions suffice for efficient computation of the optimal contract (Theorem~\ref{thm:optimal-contract}).

\subsubsection{Algorithm for Agent's Best Response} \label{sec:ultra-spa-demand}

In this section we present Algorithm~\ref{alg:ultra-spa-demand-query-contract} for efficiently computing an agent's best response in our setting.
Notably, this algorithm does not follow the \DemandForSPA\ framework (Algorithm~\ref{alg:demand-spa}), which is also polynomial-time but less efficient, and is used mainly to analyze the size of the critical values set (see Section~\ref{sec:gs-spa-critical-values} and Algorithm~\ref{alg:alternative-contract-demand-gs-spa}). The proof of Proposition~\ref{prop:contract-demand-query-ultra-add-sym} is deferred to Appendix~\ref{sec:appendix-ultra-spa}.

\begin{restatable}{proposition}
{contractdemandqueryultraaddsym}
\label{prop:contract-demand-query-ultra-add-sym}
Let $f:\setfunc$ be an \Ultra\ reward function, $c=g+p$ an \SPA\ cost function, and $\alpha \in [0,1]$ a contract.
Then Algorithm~\ref{alg:ultra-spa-demand-query-contract} computes an agent's best response; that is,\\
$\GreedyUltraSPA(\alpha, f, p, g) \in D^*_{f,c}(\alpha)$.
\end{restatable} 

\begin{proof}
Let $\Sa^*$ be the set returned by $\GreedyUltraSPA(\alpha, f, p, g)$.
To prove the lemma, we need to show that $\Sa^* \in D_{f,c}(\alpha)$, and $f(\Sa^*) \geq f(S')$ for every $S'\in D_{f,c}(\alpha)$.
The fact that $\Sa^* \in D_{f,c}(\alpha)$ follows by observing that $\GreedyUltraSPA(\alpha, f, p,g )$ is equivalent to  $\GreedyUltraVal(\alpha f-g, p)$ when the same tie breaking is used, and that if $f$ is \Ultra, and $g$ is symmetric, then $\alpha f-g$ is also \Ultra\ (the latter is stated and proved in Lemma~\ref{lemma:alpha-f-is-ultra} in Appendix~\ref{sec:appendix-ultra-add}).
The fact that $f(\Sa^*) \geq f(S')$ for every $S'\in D_{f,c}(\alpha)$ is stated and proved in Lemma~\ref{lemma:ultra-spa-maximal-f} in Appendix~\ref{sec:appendix-ultra-spa}.
\end{proof}

\begin{minipage}{0.75\linewidth}
\begin{algorithm}[H]
\caption{$\GreedyUltraSPA(\alpha, f, p, g)$}
\label{alg:ultra-spa-demand-query-contract}

\begin{algorithmic}[1]
\State Initialize $S_0, S_1, \ldots, S_n \gets \emptyset$
\For{$i = 1,\ldots,n$}
    \State \label{st:alg-ultra-spa-demand-query-contract-i}  
     Let $x_i \in \argmax_{x\notin S_{i-1}}{ \big(\alpha f(x \mid S_{i-1})-p(x) -g(x \mid S_{i-1}) \big)}$
    \State $S_i \gets S_{i-1} + x_i$
\EndFor
\State \label{st:alg-ultra-spa-demand-query-contract-final} Let $\Sa^* \in \argmax_{(S_i)_{i \in [n]}}{\big( \alpha f(S_i)-p(S_i)-g(S_i) \big)}$ 
\State \Return $\Sa^*$
 \Statex Break ties in steps (\ref{st:alg-ultra-spa-demand-query-contract-i}) and (\ref{st:alg-ultra-spa-demand-query-contract-final}) consistently and arbitrarily.

\end{algorithmic}
\end{algorithm}
\end{minipage}

\subsubsection{Poly-Many Critical Values} \label{sec:ultra-spa-critical-values}

The following theorem establishes a polynomial upper bound on the size of the critical values set.

\begin{theorem} \label{thm:ultra-add-sym-critical-values}
Given an \Ultra\ reward function $f:\setfunc$ and a \SPA\ cost function $c:\setfunc$,
\[
    |C_{f,c}| \leq \frac{n^2(n+1)(n+2)}{2}
\]
\end{theorem}

To analyze the critical values under \Ultra\ rewards and \SPA\ costs we define an alternative, less efficient best response algorithm (Algorithm~\ref{alg:alternative-contract-demand-ultra-spa}), used solely for analysis purposes.
This algorithm implements Algorithm~\ref{alg:demand-spa} (\DemandForSPA). To present Algorithm~\ref{alg:alternative-contract-demand-ultra-spa}, we first show how to compute an agent's best response when the reward is a truncated \Ultra\ function. This is cast in the following lemma, whose proof is deferred to Appendix~\ref{sec:appendix-ultra-spa}.

\begin{restatable}
{lemma}
{greedyuptot}
\label{lemma:greedy-up-to-t}
Let $f:\setfunc$ be an \Ultra\ function,
$p:\setfunc$ be an additive function, 
and $\alpha\in [0,1]$ a contract.
For any $t\in[n]$, let $f_t$ be the $t$-truncation of $f$ (See Def.~\ref{def:f_t}).
Algorithm~\ref{alg:greedy-up-to-t} computes the agent's best response for $\alpha, f_t, c$.
Namely, let $\Sa^* \gets\GreedyUpToT(\alpha, t, f, p)$. Then 
$\Sa^*\in D^*_{f_t,p}(\alpha)$ and moreover, $|\Sa^*| \leq t$.
\end{restatable}

\begin{proof}
Let $\Sa^* \gets \GreedyUpToT(\alpha, t, f, p)$.
To prove the lemma, we need to show that the following two conditions hold: $(i)$ $\Sa^* \in D_{f_t,p}(\alpha)$, and $(ii)$ $f_t(\Sa^*) \geq f_t(S')$ for every $S'\in D_{f_t,p}(\alpha)$.

To prove $(i)$, we show that $\Sa^*$ has the highest utility among any set $S \subA$. 
To this end,
let $S_1,\ldots,S_n$ denote the sequence produced by  $\GreedyUltraContract(\alpha, f, p)$. 
By construction of the algorithms, it is easy to verify that up to iteration $t$ (inclusive),  $\GreedyUpToT(\alpha, t, f, p) $ follows the same path and therefore, it produces the sequence  $S_1,\ldots,S_t$. Moreover, from the definition of $f_t$ and the \wl\ property of \Ultra\ function $f$ (Definition~\ref{def:ultra-new}), we have for all  $i\leq t$:
\[
    S_i \in 
    \argmax_{S, |S|=i} \Big(\alpha f(S)-p(S)\Big) = 
    \argmax_{S, |S|=i} \Big(\alpha f_t(S)-p(S)\Big).
\]
Furthermore, by Observation~\ref{obs:f_t-structural-insights} (\ref{en:f_t-structural-insights-3}), it holds that
    $S_t \in \argmax_{S, |S|>t} \Big(\alpha f_t(S)-p(S)\Big)$.
    
For any set $S$ with $k=|S|$ we distinguish between the following two cases:
\begin{itemize}
    \item If $k\leq t$,  then 
          $\alpha f_t(S)-p(S) \leq 
           \alpha f_t(S_k)-p(S_k)$.
    \item If $k>t$ then 
            $\alpha f_t(S)-p(S) \leq 
             \alpha f_t(S_t)-p(S_t)$. 
\end{itemize}
In both cases, $\alpha f_t(S)-p(S) \leq f_t(\Sa^*)-p(\Sa^*)$, hence, $\Sa^* \in D_{f_t,p}(\alpha)$. 
Finally, since $\Sa^*$ is selected from $\{ S_1, \ldots, S_t \}$, it satisfies $|\Sa^*|\leq t$.

The proof of condition $(ii)$ appears in Lemma~\ref{lemma:greedy-up-to-t-maximal-f} in Appendix~\ref{sec:appendix-ultra-spa}.
\end{proof}

\begin{minipage}{0.8\linewidth}
\begin{algorithm}[H]
\caption{$\GreedyUpToT(\alpha, t, f, p)$}
\label{alg:greedy-up-to-t}
\begin{algorithmic}[1]
\State Initialize $S_0, S_1, \ldots, S_t \gets \emptyset$
\For{$i = 1,\ldots,t$}
    \State \label{st:alg-greedy-up-to-t-i} Let $x_i \in \argmax_{x\notin S_{i-1}}{\big(\alpha f(x \mid S_{i-1})-p(x)\big)}$
    \State $S_i \gets S_{i-1} + x_i$
\EndFor
\State \label{st:alg-greedy-up-to-t-final} Let $\Sa^* \in \argmax_{(S_i)_{ i\in [t]}}{ \big(\alpha f(S_i)-p(S_i)\big)}$ 
\State \Return $\Sa^*$
\Statex Break ties in steps (\ref{st:alg-greedy-up-to-t-i}) and (\ref{st:alg-greedy-up-to-t-final}) consistently in favor of the larger $f$ value. 
\end{algorithmic}
\end{algorithm}
\end{minipage}

\medskip

The following corollary states that \GreedyUpToT, with suitable parameters, can be used to implement step (\ref{st:alg-demand-spa-i}) in Algorithm~\ref{alg:demand-spa} for \Ultra\ rewards.

\begin{corollary} \label{cor:ultra-demand-up-to-t}
    Let $f:\setfunc$ be an \Ultra\ reward function, $p:\setfunc$ be an additive cost function and $\alpha\in[0,1]$ be a contract. Then for every $t\in [n]$, 
    $\GreedyUpToT(\alpha, t, f, p)\in D^{\leq t}_{f,p}(\alpha)$.
\end{corollary}

\begin{proof}
Let $\Sa^*\gets \GreedyUpToT(\alpha, t, f, p)$. From Lemma~\ref{lemma:greedy-up-to-t}, $\Sa^* \in D_{f_t,p}(\alpha)$ and $|\Sa^* |\leq t$. Therefore, $\Sa^* \in D^{\leq t}_{f_t,p}(\alpha)=D^{\leq t}_{f,p}(\alpha)$, concluding the proof.
\end{proof}

The following lemma (whose proof is deferred to  Appendix~\ref{sec:appendix-ultra-spa})
will be used to analyze the size of the critical values set. 

\begin{restatable}{lemma}{greedyuptotSisubsetSj} \label{lemma:greedy-up-to-t-Si-subset-Sj}
    For $i<j$,
    $\GreedyUpToT(\alpha, i,f,p) \subseteq \GreedyUpToT(\alpha, j,f,p)$.
\end{restatable}

We next establish an upper bound on the size of the critical values set for $f_t$.

\begin{theorem} \label{thm:ultra-ft-add-critical-values}
For any \Ultra\ set function $f:\setfunc$ and any $t\in [n]$ 
we have
$|C_{f_t,p}| \leq \frac{n(n+1)}{2}$.
\end{theorem}

\begin{proof}
By Lemma~\ref{lemma:greedy-up-to-t}, Algorithm~\ref{alg:greedy-up-to-t} finds a best response set for $f_t$ and our bound of the size of the set of critical values with respect to $f_t$ relies on the analysis of Algorithm~\ref{alg:greedy-up-to-t}. 
This algorithm is identical to Algorithm~\ref{alg:ultra-demand-query-contract}, except that it iterates only up to $t \leq n$, whereas Algorithm~\ref{alg:ultra-demand-query-contract} always iterates up to $n$. The proof of Theorem~\ref{thm:ultra-add-critical-values} is based on the analysis of Algorithm~\ref{alg:ultra-demand-query-contract}. 
It is not too difficult to observe that the same analysis as in the proof of 
Theorem~\ref{thm:ultra-add-critical-values} carries over to the analysis of Algorithm~\ref{alg:greedy-up-to-t}, simply by restricting all iterations to go only up to $t$, in all arguments along the proof.
\end{proof}

\begin{restatable}
{lemma}
{altgreedyultraspa}
\label{lemma:alt-greedy-ultra-spa}

For an \Ultra\ reward function $f$, a \SPA\ cost function $c=g+p$, and a contract $\alpha\in[0,1]$, Algorithm~\ref{alg:alternative-contract-demand-ultra-spa} computes an agent's best response. Namely, \\
$\AltGreedyUltraSPA(\alpha, f, p, g)\in D^*_{f,c}(\alpha)$.
\end{restatable}

\begin{proof}
Observe that Algorithms~\ref{alg:alternative-contract-demand-ultra-spa} and~\ref{alg:demand-spa} differ only in step~(\ref{st:alg-demand-spa-i}).
To show that $\Sa^* \in D^*_{f,c}(\alpha)$ it suffices to prove that the following two conditions hold for every $i \in [n]$: $(i)$ $S_i \in D^{\leq i}_{f,p}(\alpha)$, and $(ii)$ there exists some $\epsilon_i>0$, such that  $\GreedyUpToT((\alpha+\epsilon), f, p)=\GreedyUpToT(\alpha, f, p)$ for all $0<\epsilon\leq \epsilon_i$.
Condition (i) follows from Corollary~\ref{cor:ultra-demand-up-to-t}, and condition (ii) follows from the structure of \GreedyUpToT\ and is shown in Lemma~\ref{lemma:greedy-up-to-t-alpha-plus-epsilon} (Appendix~\ref{sec:appendix-ultra-spa}).
By Lemma~\ref{lemma:demand-for-spa}, we conclude that $\Sa^* \in D^*_{f,c}(\alpha)$.
\end{proof}

\begin{minipage}{0.8\linewidth}
\begin{algorithm}[H]
\caption{\AltGreedyUltraSPA($\alpha$, f, p, g)}
\label{alg:alternative-contract-demand-ultra-spa}
\begin{algorithmic}[1]
\For{$i = 1,\ldots,n$}
    \State \label{st:alg-alt-ultra-spa-i} $S_i \gets \GreedyUpToT(\alpha , i, f,p)$ \Comment{ $S_i \in D^{\leq i}_{f,p}(\alpha)$}
\EndFor

\State \label{st:alg-alt-ultra-spa-final} Let $\Sa \in \argmax_{(S_i)_{i \in [n]}} \big( \alpha
 f(S_i)-p(S_i)-g(S_i) \big) $ 
\State \Return $\Sa^*$
\Statex Break ties in steps (\ref{st:alg-alt-ultra-spa-i}) and (\ref{st:alg-alt-ultra-spa-final}) consistently in favor of the larger $f$ value.
\end{algorithmic}
\end{algorithm}
\end{minipage}





\medskip

The proof of Lemma~\ref{lemma:alt-greedy-ultra-spa} appears in Appendix~\ref{sec:appendix-ultra-spa}.

We are now ready to prove Theorem~\ref{thm:ultra-add-sym-critical-values} by analyzing \AltGreedyUltraSPA. Namely, an upper bound of $n^2(n+1)(n+2)/2$ on the size of the set of critical values for \Ultra\ rewards and \SPA\ costs.

\begin{proof}[Proof of Theorem~\ref{thm:ultra-add-sym-critical-values}]
Let $\bigcup_{1\leq i\leq n} C_{f_i, p} = 
    \{ \alpha_0=0,\alpha_1,\ldots \alpha_k \}$, ordered by $\alpha_i<\alpha_j$ for every $i<j$.
From Theorem~\ref{thm:ultra-ft-add-critical-values}, we have
\[
    k = 
    |\bigcup_{1\leq i\leq n} C_{f_i, p}| \leq 
    \sum_{1\leq i\leq n} |C_{f_i, p}| \leq 
    n \cdot\frac{n(n+1)}{2} = 
    \frac{n^2(n+1)}{2}
\] 

For every $1\leq \ell \leq k$, define $B_\ell := C_{f,c} \cap (\alpha_{\ell-1},\alpha_{\ell})$ to be the critical values from $C_{f,c}$ that are in the interval $(\alpha_{\ell-1},\alpha_{\ell})$. 
Notice that
\[
    C_{f,c} \subseteq 
    \bigcup_{1\leq \ell \leq k} B_\ell +
    \{\alpha_0,\alpha_1,\ldots,\alpha_k\} .
\]
Next, we show that there are at most $n+1$ critical values in each $B_\ell$. Let there be $1 \leq \ell \leq k$ and some $\beta\in B_\ell$. 
Observe that $\beta \notin \bigcup_{1\leq i\leq n} C_{f_i, p}$. 
For every $i$, let $S_i^{\beta} \gets \GreedyUpToT(\beta, i, f, p)$ denote the sets produced by running $\AltGreedyUltraSPA(\beta, f, p, g)$ in iteration $i$, and denote its result by $S_\beta^*$.  
Let us also denote the result of the algorithm on contract $(\beta-\epsilon)$, for some $\epsilon$, by $S_{\beta-\epsilon}^*$.

First, we show that there exists some $\epsilon'>0$, for which running \AltGreedyUltraSPA\ on contract $(\beta-\epsilon')$ produces the same sequence $S_1^\beta, \ldots, S_n^\beta$. 
To this end, since $\beta \notin \bigcup_{1\leq i\leq n} C_{f_i, p}$, then for every $i$, $\beta \notin C_{f_i, p}$, so there must exist some $\epsilon_i>0$, for which $\GreedyUpToT(\beta, i, f, p)$ returns $S_i$ as well. 
Clearly, for $\epsilon':=\min \{\epsilon_1, \ldots, \epsilon_n \}$, for every $i$, $\GreedyUpToT((\beta-\epsilon'), i, f, p)$ returns $S_i$ as well so the claim holds. 

Next, since $\beta \in C_{f,c}$, there must exist some sufficiently small $\epsilon''>0$, for which running algorithm \AltGreedyUltraSPA\ on contracts $\beta$ and $(\beta-\epsilon'')$ yields different results $S_\beta^* \neq S_{\beta-\epsilon''}^*$. 
Therefore, it is clear that for $\epsilon:=\min \{\epsilon', \epsilon''\}$, running \AltGreedyUltraSPA\ on contracts $\beta$ and $(\beta-\epsilon)$ produce the same sequence $S_1^\beta, \ldots, S_n^\beta$, but then yield different results $S_\beta^* \neq S_{\beta-\epsilon}^*$.
These different results are chosen in step (\ref{st:alg-alternative-gs-spa-final}).
%
Let $i,j$ be such that $S_{\beta-\epsilon}^* = S_i^\beta$ and $S_{\beta}^* =S_j^\beta$. By the monotonicity lemma (Lemma~\ref{lemma:monotonicity-of-critical-values}) we have $f(S_{\beta-\epsilon}^*) < f(S_{\beta}^*)$ so $f(S_i^\beta) < f(S_j^\beta)$. 
Moreover, by Lemma~\ref{lemma:greedy-up-to-t-Si-subset-Sj} and the monotonicity of $f$, it must be that $i < j$ and thus $S_i^\beta \subset S_j^\beta$.
Hence, $S_{\beta-\epsilon}^* \subset S_{\beta}^*$.

Now let $B_\ell=\{ \beta_1, \ldots, \beta_r\}$ with $r=|B_\ell|$. 
Then $S_{\beta_1}\subset \ldots S_{\beta_r}\subseteq A$, and 
therefore $r\leq n+1$. 
To conclude, $|B_\ell|\leq n+1$ for every $1\leq \ell\leq n $.
We get
\[
    |C_{f,c}| \leq 
    | \bigcup_{1\leq \ell \leq k} B_\ell | + 
    |\{\alpha_0,\alpha_1,\ldots,\alpha_k\} | =
    k(n+1) + k \leq
    \frac{n^2(n+1)(n+2)}{2}.\]
\end{proof}

\bibliographystyle{alpha}
\bibliography{bib}

\appendix

\section{Optimal Contracts for Weakly Well-Layered (WWL) Rewards and Symmetric Costs} 
\label{sec:wwl} \label{sec:appendix-wwl-1}

In this section we present an algorithm for computing the optimal contract in instances with \wwl\ rewards and monotone symmetric costs. \WWL\ is a seuperset of \wl.

Interestingly, \cite{efx-wwl} show that the class of {\em budget-additive} functions is a subclass of \wwl. As a direct corollary, instances with budget-additive rewards and symmetric costs admit a polynomial-time algorithm for the optimal contract problem.
This is particularly interesting given the \textsf{NP}-hardness result of \cite{DEFK21} for budget-additive rewards with {\em additive} costs.

We start by defining the class of \wwl\ functions.

\begin{definition}[WWL]\cite{efx-wwl}
\label{def:wwl}
Let $f:\setfunc$ be a set function. Consider the sequence of sets 
$S_0 \subseteq S_1 \subseteq \ldots \subseteq S_n$, with $S_0=\emptyset$, 
constructed by the following greedy procedure.

For $i=1,\ldots,n$, let $S_i = S_{i-1} + x_i$, where 
$x_i \in \argmax_{x\notin S_{i-1}} f(x \mid S_{i-1})$, with ties broken arbitrarily.

We say that $f$ is \Wwl\ (\WWL) if for every $i \in [n]$, $S_i \in \argmax_{S:|S|=i} f(S)$.
\end{definition}

Note that this definition is weaker than that of \wl\ functions. It is based on a greedily optimizing the marginal value only with respect to 
$f$ at each step, without accounting for prices.

\begin{proposition}
\label{prop:optimal-contract-wwl-sym}
Let $f$ be a \WWL\ reward function and $c$ a symmetric monotone cost function (i.e., depending only on the set size). Then the optimal contract problem can be solved in polynomial time.
\end{proposition}

Recall that by Theorem~\ref{thm:optimal-contract}, a sufficient condition for a poly-time algorithm for the optimal contract problem is a poly-time algorithm for computing an agent's best response and poly-many critical values. 
Any setting with symmetric costs clearly has at most $n+1$ critical values. Thus, it remains to show an algorithm for computing an agent's best response. 
This is shown in the following lemma.


\begin{lemma}
Let $f:\setfunc$ be a \wwl\ reward function, $g:\setfunc$ a symmetric monotone cost function, and $\alpha \in [0,1]$. 
Then Algorithm~\ref{alg:wwl-contract-demand-query} computes an agent's best response; that is,
$\GreedyWWL(\alpha, f, g) \in D^*_{f,g}(\alpha)$.
\end{lemma}


\begin{minipage}{0.75\linewidth}
\begin{algorithm}[H]
\caption{$\GreedyWWL(\alpha, f, g)$:}
\label{alg:wwl-contract-demand-query}
\begin{algorithmic}[1]
\State Initialize $S_0, S_1, \ldots, S_n \gets \emptyset$
\For{$i = 1,\ldots,n$}
    \State \label{st:alg-wwl-i} Choose $x_i \in \argmax_{x\notin S_{i-1}}{ f(x \mid S_{i-1})}$
    \State $S_i \gets S_{i-1} + x_i$
\EndFor
\State \label{st:alg-wwl-final} Choose $S_\alpha^* \in \argmax_{(S_i)_{i \in [n]}}{ \big(f(S_i)-g(S_i)\big)}$
\State \Return $S_\alpha^*$
\Statex Break ties in steps (\ref{st:alg-wwl-i}) and (\ref{st:alg-wwl-final}) consistently in favor of larger $f$ value. 

\end{algorithmic}
\end{algorithm}
\end{minipage}

    


\begin{proof}
Let $\Sa^* \gets \GreedyWWL(\alpha, f, g)$.
To prove the lemma, we need to show that the following two conditions hold: $(i)$ $\Sa^* \in D_{f,g}(\alpha)$, and $(ii)$ $f(\Sa^*) \geq f(S')$ for every $S'\in D_{f,g}(\alpha)$.

To establish $(i)$, we show that $\Sa^*$ has the highest utility out of any set $S \subA$. To this end, observe that by the definition of \WWL, $S_i \in \argmax_{S,|S|=i} f(S)$ for every $i$. Since $g$ is symmetric, it assigns the same value to all sets of size $i$, so 
$S_i \in \argmax_{S,|S|=i} f(S)-g(S)$ as well.
Therefore, 
\[\max_{S\subA } \Big(\alpha f(S)-g(S) \Big)= 
 \max_{i \in [n]} \max_{S, |S|=i} \Big(\alpha f(S)-g(S)\Big) = 
 \max_{i \in [n]} \Big( \alpha f(S_i)-g(S_i) \Big) = 
 \alpha f(\Sa^*)-g(\Sa^*).\]

To prove $(ii)$, first observe that the choice of $S_1,\ldots,S_n$ is independent of $\alpha$. Moreover, there must exist a sufficiently small $\epsilon>0$, for which, for every $i$ with $S_i \neq S_\alpha^*$ and $\alpha f(\Sa^*)-g(S_\alpha^*) > \alpha f(S_i)-g(S_i)$, 
it must also hold that 
$(\alpha+\epsilon) f(S_\alpha^*)-g(S_\alpha^*) > 
 (\alpha+\epsilon) f(S_i)-g(S_i)$. 
Furthermore, every $i$ with $S_i \neq S_\alpha^*$ and $\alpha f(\Sa^*)-g(S_\alpha^*) = \alpha f(S_i)-g(S_i)$, by the tie-breaking rule, has $f(\Sa^*) \geq f(S_i)$, and therefore, 
$(\alpha+\epsilon) f(S_\alpha^*)-g(S_\alpha^*) \geq 
 (\alpha+\epsilon) f(S_i)-g(S_i)$.
Consequently, executing \GreedyWWL\ on contracts $\alpha$ and $(\alpha+\epsilon)$ would follow the same path and return the same result $\Sa^*$, therefore $S_\alpha^* \in D_{f,g}(\alpha+\epsilon)$.
Thus, by the monotonicity lemma (Lemma~\ref{lemma:monotonicity-of-critical-values}), $f(S_{\alpha}^*) \geq f(S')$ for any $S'\in D_{f,g}(\alpha)$. 
\end{proof}


\section{\GS\ Rewards:  Beyond Additive Costs} \label{sec:appendix-gs-beyond-additive}

In this section, we prove a variant of Theorem~\ref{thm:ultra-spa} tailored to the special case where the reward function is \GS\ (see Theorem~\ref{thm:gs-spa-optimal-contract}).
While the general proof for \Ultra\ rewards also applies to the \GS\ case, we present a separate argument that is considerably simpler and more direct in this setting.

\begin{theorem} \label{thm:gs-spa-optimal-contract}
Given a \GS\ reward function and a \SPA\ cost function, the optimal contract can be computed in polynomial time. 
\end{theorem}

We provide a poly-time algorithm for computing an agent's best response in this setting in Section~\ref{sec:gs-spa-demand}, and we show that there are poly-many critical values in section~\ref{sec:gs-spa-critical-values}.
By Theorem~\ref{thm:optimal-contract}, these two conditions suffice to solve the optimal contract problem efficiently.

\subsection{Algorithm for Best Response} \label{sec:gs-spa-demand}

We present Algorithm~\ref{alg:contract-demand-query-gs-spa} that computes an agent's best response for \GS\ rewards and \SPA\ costs.
Although Algorithm~\ref{alg:ultra-spa-demand-query-contract} (best response oracle for \Ultra\ rewards and \SPA\ costs) is applicable in this setting as well, since \GS\ functions form a subclass of \Ultra, we propose a different algorithm that is slightly more efficient.
Importantly, the structural distinction between Algorithms~\ref{alg:contract-demand-query-gs-spa} and~\ref{alg:ultra-spa-demand-query-contract} mirrors the distinction between Algorithms~\ref{alg:gs-demand-query} and~\ref{alg:ultra-demand-query} (demand query algorithm for \GS\ and \Ultra\ valuations, respectively).

\begin{lemma} \label{lemma:contract-demand-query-gs-f-add-sym-c}
    Given a \GS\ reward function $f:\setfunc$, a \SPA\ cost function $c=g+p$, and a contract $\alpha\in [0,1]$, 
    Algorithm~\ref{alg:contract-demand-query-gs-spa} computes an agent's best response. Namely, \\$\GreedyGSSPA(\alpha,f,p,g) \in D^*_{f,c}(\alpha)$.
\end{lemma}


\begin{minipage}{0.8\linewidth}
\begin{algorithm}[H]
\caption{$\GreedyGSSPA(\alpha, f, p, g)$}
\label{alg:contract-demand-query-gs-spa}
\begin{algorithmic}[1]
\State Initialize $S_0, S_1, \ldots, S_n \gets \emptyset$
\For{$i = 1,\ldots,n$}
    \State \label{st:alg-gs-spa-i} Let $x_i \in \argmax_{x\notin S_{i-1}}{\big( \alpha f(x \mid S_{i-1})-p(x) \big)}$
    \If {$\alpha f(x_i \mid S_{i-1})-p(x_i) > 0$}
        \State $S_i \gets S_{i-1} + x_i$
    \Else 
        \State $\label{st:alg-gs-spa-break}$ Go to step~\ref{st:alg-gs-spa-final}
    \EndIf
\EndFor
\State \label{st:alg-gs-spa-final} Let $S_\alpha ^* \in \argmax_{(S_i)_{i \in [n]}}{\big( \alpha f(S_i)-p(S_i)-g(S_i) \big)}$ 
\State \label{st:alg-gs-spa-return}\label{st:alg-gs-spa-last} \Return $S_\alpha^*$
\Statex Break ties in steps (\ref{st:alg-gs-spa-i}) and (\ref{st:alg-gs-spa-final}) consistently in favor of the larger $f$ value. 
\end{algorithmic}
\end{algorithm}
\end{minipage}

    



\medskip

\noindent
The correctness of Algorithm~\ref{alg:contract-demand-query-gs-spa} relies on the following observation. For every $i$, let $g_i$ denote the value that $g$ obtains on a set of size $i$, then
\[
\max_{S, |S|=i} \Big(\alpha f(S)-p(S)-g(S)\Big) = 
\max_{S, |S|=i} \Big(\alpha f(S)-p(S)\Big) - g_i
\]
That is, due to the symmetric structure of $g$, one can fix the size of $S$, compute a best $i$-sized response, while ignoring the symmetric component of the cost (an operation that can be performed efficiently for \GS\ rewards as it has the \wl\ property) and then search for the best response across the values of $i$. Formally,

\begin{proof}
Let $S_1, \ldots, S_n$ denote the sequence produced by $\GreedyGSSPA(\alpha, f, p, g)$, and let $\Sa^*$ denote its result. 

To prove the lemma, we need to show that the following two conditions hold: $(i)$ $\Sa^* \in D_{f,c}(\alpha)$, and $(ii)$ $f(\Sa^*) \geq f(S')$ for every $S'\in D_{f,c}(\alpha)$.

To prove $(i)$, we show that $\Sa^*$ achieves the highest utility among all sets $S \subA$. 
Let $g_i$ denote the value of $g$ on any set of size $i$, and let $j$ be the iteration in which step~(\ref{st:alg-gs-spa-break}) was triggered (i.e. $\alpha f(x_j | S_{j-1}) -p(x_j) \leq 0$). Thus, for every $i < j$, the set $S_i$ was updated during execution.

Note that $S_{j-1} \in D_{f,p}(\alpha)$, since if we replace steps (\ref{st:alg-gs-spa-break})-(\ref{st:alg-gs-spa-last}) by returning $S_{i-1}$, produces the same algorithm as Algorithm~\ref{alg:gs-demand-query-contract} (the best response algorithm for \GS\ rewards and additive costs).

Let there be some set $S$ with $|S|=k$. 
If $k<j$, then since every \GS\ function has the \wl\ property (Theorem~\ref{thm:gs-is-wl}), it follows that $S_k \in \argmax_{S,|S|=k} \alpha f(S)-p(S)$. Moreover, by symmetry, $g(S)=g(S_k)=g_k$, thus $\alpha f(S)-p(S)-g(S) = \alpha f(S)-p(S)-g_k \leq \alpha f(S_k)-p(S_k)-g_k = \alpha f(S_k)-p(S_k)-g(S_k)$. 
If $k \geq j$, since $S_{j-1} \in D_{f,p}$, we have $\alpha f(S)-p(S) \leq \alpha f(S_{j-1})-p(S_{j-1})$, and from monotonicity of $g$ we have $g(S)=g_k\geq g_{j-1}=g(S_{j-1})$, and therefore $\alpha f(S)-p(S)-g(S) = \alpha f(S)-p(S)-g_k \leq \alpha f(S_{j-1})-p(S_{j-1})-g_{j-1} = \alpha f(S_{j-1})-p(S_{j-1})-g(S_{j-1})$.
In either case, $\alpha f(S)-p(S)-g(S)
\leq \alpha f(S_\alpha^*)-p(S_\alpha^*)-g(S_\alpha^*)$ as needed.

To show that $(ii)$ holds, that is, $f(\Sa^*) \geq f(S')$ for every $S'\in D_{f,c}(\alpha)$, we prove that there exists some $\epsilon>0$, for which $\Sa^*\in D_{f,c}(\alpha+\epsilon)$ also. The claim then follows from the monotonicity lemma. 
To this end, in step (\ref{st:alg-gs-spa-i}) when choosing $x_i$ for every $i$, for every $x\notin S_{i-1}$ with $x\neq x_i$, one of the following holds:
\begin{enumerate}[itemsep=1pt]
    \item \label{en:alg-gs-spa-no-tiebreak-i}$\alpha f(x_i \mid S_{i-1})-p(x_i) > \alpha f(x \mid S_{i-1})-p(x)$. Or
    \item \label{en:alg-gs-spa-tiebreak-i}$\alpha f(x_i \mid S_{i-1})-p(x_i) = \alpha f(x \mid S_{i-1})-p(x)$. Then by the tie-breaking rule, $f(x_i \mid S_{i-1}) \geq f(x \mid S_{i-1})$.
\end{enumerate}
Similarly, in the final step (\ref{st:alg-gs-spa-final}),  one of the following holds for every $i$ s.t. $S_i \neq S_\alpha^*$,
\begin{enumerate}[itemsep=1pt]
    \setcounter{enumi}{2}  
    \item \label{en:alg-gs-spa-no-tiebreak-final} $\alpha f(\Sa^*)-p(\Sa^*)-g(\Sa^*) > \alpha f(S_i)-p(S_i)-g(S_i)$. Or
    \item \label{en:alg-gs-spa-tiebreak-final} $\alpha f(S^*)-p(\Sa^*)-g(\Sa^*) = \alpha f(S_i)-p(S_i)-g(S_i)$. Then by the tie-breaking rule, $f(\Sa^*) \geq f(S_i)$.
\end{enumerate}
Clearly, there exists a sufficiently small $\epsilon>0$ such that, for every $i$ and for every $x\notin S_{i-1}$ with $x\neq x_i$ satisfying (\ref{en:alg-gs-spa-no-tiebreak-i}), it also holds that
\[
    (\alpha+\epsilon) f(x_i \mid S_{i-1})-p(x_i) > 
    (\alpha+\epsilon) f(x \mid S_{i-1})-p(x)
\]
and for every $i$ with $S_i\neq \Sa^*$, satisfying (\ref{en:alg-gs-spa-no-tiebreak-final}), we also have
\[
    (\alpha+\epsilon) f(\Sa^*)-p(\Sa^*)-g(\Sa^*) > 
    (\alpha+\epsilon) f(S_i)-p(S_i)-g(S_i)
\]
Moreover, for every $i$ and every $x\notin S_{i-1}$ with $x\neq x_i$, satisfying (\ref{en:alg-gs-spa-tiebreak-i}), it also holds that
\[
    (\alpha+\epsilon) f(x_i \mid S_{i-1})-p(x_i) \geq 
    (\alpha+\epsilon) f(x \mid S_{i-1})-p(x),
\]
and for every $i$ with $S_i \neq \Sa^*$ satisfying (\ref{en:alg-gs-spa-tiebreak-final}), it also holds that
\[
    (\alpha+\epsilon) f(\Sa^*)-p(\Sa^*)-g(\Sa^*) \geq 
    (\alpha+\epsilon) f(S_i)-p(S_i)-g(S_i)
\]
Consequently, executing \GreedyGSSPA\ on contracts $\alpha$ and $(\alpha+\epsilon)$ would follow the same path and return the same result $\Sa^*$, therefore $\Sa^*\in D_{f,c}(\alpha+\epsilon)$. Thus, by the monotonicity lemma (Lemma~\ref{lemma:monotonicity-of-critical-values}), every $S'\in D_{f,c}(\alpha)$ has $f(\Sa^*)\geq f(S')$.
\end{proof}

\subsection{Poly-Many Critical Values} \label{sec:gs-spa-critical-values}
The following theorem establishes a polynomial upper bound on the size of the critical values set.

\begin{theorem} \label{thm:gs-spa-critical-values}
    Given a \GS\ reward function $f:\setfunc$ and a \SPA\ cost function $c:\setfunc$,  
\[
    |C_{f,c}| \leq \frac{n^2(n+1)(n+2)}{2}
\]
\end{theorem}

To analyze the critical values under \GS\ rewards and \SPA\ costs, we introduce an alternative, less efficient algorithm for computing best responses (Algorithm~\ref{alg:alternative-contract-demand-gs-spa}), used solely for this analysis.
This algorithm is an implementation of Algorithm~\ref{alg:demand-spa} (\DemandForSPA), where the key component is computing an agent's best response under truncated \GS\ reward functions.

This approach is similar to the one used to bound the critical values for \Ultra\ rewards (see Section~\ref{sec:ultra-spa-critical-values}). The difference here is the approach used to compute a best response under truncated rewards. 
It turns out that the \GS\ property is preserved under truncation, enabling the use of the \GreedyGSContract\ algorithm, assuming value oracle access to the truncated reward function is available. Since this algorithm is used only for analysis, efficiency is not a concern.

We begin by describing how to compute an agent's best response for a truncated \GS\ function, and then present the full algorithm. But before getting to that, we need to make a general observation about \GS\ functions that will be used in the following proofs. 

\begin{observation} \label{obs:gs_arg_max_leq_i}
For any \GS\ set function $f:\setfunc$, let $S_1, \ldots, S_n$ denote the sets produced by running $\GreedyGSVal(f, p)$, and let $S^*$ denote its result, with $|S^*|=k$.
Then for every $i<k$, $S_i \in \argmax_{S, |S|\leq i} \Big(f(S)-p(S)\Big)$ .
\end{observation}
\begin{proof}
Observe that by definition of the \GreedyGSVal, $S^*=S_k$ (the algorithm stops after $k$ iterations and returns $S_k$). 
Moreover, for every $i\leq k$ it holds that  $f(x_{i} \mid S_{i-1}) - p(x_{i}) > 0$ (since the algorithm proceeds to run only if this condition holds). 
Rearranging, we get 
\begin{align*}
    f(x_{i} \mid S_{i-1}) - p(x_{i}) = 
    & f(x_{i} + S_{i-1}) - f(S_{i-1}) - \Big( p(x_{i} + S_{i-1}) - p(S_{i-1}) \Big) \\
    = & f(S_{i}) - p(S_{i}) - \Big( f(S_{i-1}) - p(S_{i-1}) \Big) > 0
\end{align*}

So $f(S_{i-1})-p(S_{i-1})<f(S_{i})-p(S_{i})$ for every $i \leq k$, and inductively, $f(S_{i})-p(S_{i})<f(S_{k})-p(S_{k})$ as well. 
Since \GS\ has the  \wl\ property (Theorem~\ref{thm:gs-is-wl}), it follows that for every $i\leq k$, $S_i \in \argmax_{S,|S|=i} f(S)-p(S)$, and altogether, $S_i \in \argmax_{S,|S|\leq i} f(S)-p(S)$ as well.
\end{proof}

We are now ready to prove that truncation preserves \GS. 

\begin{theorem} \label{thm:f_t-is-gs} 
If $f$ is \GS, then $f_t$ is \GS\ for every $t \in [n]$. As a result, 
    \begin{enumerate}[leftmargin=15pt, itemsep=0pt]
        \item $\GreedyGSContract(\alpha, f_t,p) \in D_{f_t,p}(\alpha)$.
        \item For any additive cost function $p:\setfunc$, $|C_{f_t,p}| \leq \frac{n(n+1)}{2}$.
    \end{enumerate}
\end{theorem}

\begin{proof}
As a result from \cite[Theorem 3.2]{gs-survey}, 
it suffices to show that $f_t$ is "greedily solvable", that is, for any additive function $p:\setfunc$, $\GreedyGSVal(f_t, p) \in D_{f_t}(p)$. 

Let $S_1,\ldots, S_n$
denote the sequence of sets produced by running $\GreedyGSVal(f, p)$, and let $S^*$ denote its result (thus, $S^* \in D_{f}(p)$). 
By the definition of $f_t$, it is easy to see that up to iteration $t$ (inclusive), the computations made by \GreedyGSVal on $f$ and $f_t$ are the same. Thus, the first $t$ iterations of $\GreedyGSVal(f_t, p)$ produce the sequence $S_1, \ldots, S_t$, and the algorithm proceeds from there.

Let $k$ be such that $S^*=S_k$, (that is, $\GreedyGSVal(f, p)$ stops after $k$ iterations and returns $S_k$). Thus, $S_k \in D_f(p)$. 
Moreover, if $k\leq t$, from Observation~\ref{obs:f_t-structural-insights} (\ref{en:f_t-structural-insights-2}), $S^* \in D_{f_t}(p)$ as well. Furthermore, it is easy to validate that $\GreedyGSVal(f_t, p)$ stops after $k$ iterations, and returns $S_k=S^*$ as well in this case.

Assume that $k> t$. We show that in this case, $\GreedyGSVal(f_t, p)$ stops after $t$ iterations and returns $S_t$.  
To this end, 
denote  the choice of $\GreedyGSVal(f_t, p)$  in step (\ref{st:alg-gs-demand-query-i}) of Algorithm~\ref{alg:gs-demand-query} on iteration $i=t+1$ by $x^{(t)}_{t+1}$. It holds that $x^{(t)}_{t+1}$ has a non positive marginal utility, since
\[
    f_t(x^{(t)}_{t+1} \mid S_{t})-p(x^{(t)}_{t+1}) = 
    f_t(x^{(t)}_{t+1}+S_{t}) -f_t(S_{t})-\Big(p(x^{(t)}_{t+1}+S_{t}) - p(S_{t})\Big) =
    \]\[
    \Big(f_t(x^{(t)}_{t+1}+S_{t}) -  p(x^{(t)}_{t+1}+S_{t})\Big) - \Big(f_t(S_{t}) - p(S_{t})\Big) 
    \leq 0.
\]
Therefore $\GreedyGSVal(f_t, p)$  stops and returns $S_t$. It is left to show that $S_t \in D_{f_t}(p)$. To this end, by Observation~\ref{obs:gs_arg_max_leq_i}, $S_t \in \argmax_{S,|S|\leq t} \Big( f(S)-p(S) \Big)$ (since $k>t$), and therefore, by Observation~\ref{obs:f_t-structural-insights} (\ref{en:f_t-structural-insights-1}), $S_t \in D_{f_t}(p)$ as needed. As a result, $f_t$ is "greedy solvable".
\end{proof}

The following claim states that \GreedyGSContract, with suitable parameters can be used to implement step (\ref{st:alg-demand-spa-i}) in Algorithm~\ref{alg:demand-spa} for \GS\ rewards. It is a corollary of the proof of Theorem~\ref{thm:f_t-is-gs}. 

\begin{corollary} \label{cor:result-of-greedy-gs-f_t}
Let $\Sa^* \gets \GreedyGSContract(\alpha,f_t, p)$ for som $t \in [n]$. Then $|\Sa^*| \leq t$ and $\Sa^* \in D^{\leq t}_{f,p}(\alpha)$.
\end{corollary}
\begin{proof}
In the proof of Theorem~\ref{thm:f_t-is-gs} it was shown that \GreedyGSContract\ never proceeds after iteration $t$, therefore $|\Sa^*| \leq t$.
Moreover, by Theorem~\ref{thm:f_t-is-gs},  $\Sa^* \in D_{f_t,p}(\alpha)$. Thus, $\Sa^* \in D^{\leq t}_{f_t,p}(\alpha)=D^{\leq t}_{f,p}(\alpha)$.
\end{proof}

The following observation is used to analyze the size of the set of critical values. 

\begin{observation}
\label{obs:result-of-greedy-gs-f_i-f_j}
Let $i<j$, $S_i \gets \GreedyGSContract(\alpha, f_i, p)$ and $S_j \gets \GreedyGSContract(\alpha, f_j, p)$. Then $S_i \subseteq S_j$.
\end{observation}
\begin{proof}
Let $S_i=\{x^{(i)}_1, \ldots, x^{(i)}_{i'}\}$ and $S_j=\{x^{(j)}_1, \ldots, x^{(j)}_{j'}\}$ for some $i', j'$. 
By Corollary~\ref{cor:result-of-greedy-gs-f_t},  $i'\leq i$ and  $j'\leq j$. 
Observe that up to iteration $i$, $\GreedyGSContract(\alpha, f_j,p)$ and $\GreedyGSContract(\alpha, f_i,p)$ make exactly the same computations (due to the consistent tie breaking) by the definition of $f_i$ and $f_j$. As a result, $x^{(j)}_k = x^{(i)}_k$ for every $k\leq i'$. 
Thus, $S_i \subseteq S_j$.
\end{proof}



We are now ready to proceed to analyzing the size of set of the critical values. We present the alternative best response algorithm, which computes best responses under the truncated reward function.


\begin{lemma}
For a \GS\ reward function $f$, a \SPA\ cost function $c=g+p$, and a contract $\alpha\in[0,1]$, Algorithm~\ref{alg:alternative-contract-demand-gs-spa} computes an agent's best response. Namely, \\$\AltGreedyGSSPA(\alpha, f, p, g)\in D^*_{f,c}(\alpha)$.
\end{lemma}


\begin{minipage}{0.8\linewidth}
\begin{algorithm}[H]
\caption{\AltGreedyGSSPA($\alpha$, f, p, g)}
\label{alg:alternative-contract-demand-gs-spa}
\begin{algorithmic}[1]
\State Initialize $S_0, S_1, \ldots, S_n \gets \emptyset$
\For{$i = 1,\ldots,n$}
    \State \label{st:alg-alternative-gs-spa-i} $S_i \gets \GreedyGSContract(\alpha, f_i, p)$ \Comment{ $S_i \in D^{\leq i}_{f,p}(\alpha)$}
\EndFor
\State \label{st:alg-alternative-gs-spa-final} Let $\Sa ^* \in \argmax_{(S_i)_{i \in [n]}}{\big(\alpha f(S_i)-p(S_i)-g(S_i)\big)}$ 
\State \Return $\Sa^*$

\Statex  Break ties in steps (\ref{st:alg-alternative-gs-spa-i}) and (\ref{st:alg-alternative-gs-spa-final}) consistently in favor of the larger $f$ value.
\end{algorithmic}
\end{algorithm}
\end{minipage}




\begin{proof}
Observe that other than step (\ref{st:alg-demand-spa-i}), Algorithms~\ref{alg:alternative-contract-demand-gs-spa} and~\ref{alg:demand-spa} are the same. Moreover, by  Corollary~\ref{cor:result-of-greedy-gs-f_t}, $S_i\in D^{\leq i}_{f,p}(\alpha)$ for every $i\in [n]$. Therefore, to show that $\Sa^* \in D^*_{f,c}(\alpha)$, 
we need to prove that the assumption from Lemma~\ref{lemma:demand-for-spa} holds, that is, for every $i \in [n]$ there exists some $\epsilon_i>0$, such that for every $0<\epsilon\leq \epsilon_i$, 
$\GreedyGSContract((\alpha+\epsilon), f, p)=\GreedyGSContract(\alpha, f, p)$. Therefore, the claim follows from Lemma~\ref{lemma:demand-for-spa}. To this end, we will show that the condition holds as a result of the structure of \GreedyGSContract. It is easy to see that there must exist some $\epsilon'>0$ such that for every $0<\epsilon\leq \epsilon'$, \GreedyGSContract\ returns the same result for $\alpha$ and for $(\alpha+\epsilon)$. Therefore, by Lemma~\ref{lemma:demand-for-spa}, $\Sa^* \in D^*_{f,c}(\alpha)$.
\end{proof}

We are now ready to prove Theorem~\ref{thm:gs-spa-critical-values} by analyzing \AltGreedyGSSPA. Namely, an upper bound of $n^2(n+1)(n+2)/2$ on the size of the set of critical values 
for \GS\ rewards and \SPA\ costs.

\begin{proof}[Proof of Theorem~\ref{thm:gs-spa-critical-values}]
Let $\bigcup_{1\leq i\leq n} C_{f_i, p} = 
    \{ \alpha_0=0,\alpha_1,\ldots \alpha_k \}$, ordered by $\alpha_i<\alpha_j$ for every $i<j$.
For every $i\in [n]$, as established in Theorem~\ref{thm:f_t-is-gs}, $f_i$ is \GS. Thus, by a result from \cite[Theorem 4.6]{DEFK21}, $|C_{f_i, p}| \leq n(n+1)/2$. Altogether, 
\[
    k = 
    |\bigcup_{1\leq i\leq n} C_{f_i, p}| \leq 
    \sum_{1\leq i\leq n} |C_{f_i, p}| \leq 
    n \cdot\frac{n(n+1)}{2} = 
    \frac{n^2(n+1)}{2}.
\] 
For every $1\leq \ell \leq k$, define $B_\ell := C_{f,c} \cap (\alpha_{\ell-1},\alpha_{\ell})$ to be the critical values from $C_{f,c}$ that are in the interval $(\alpha_{\ell-1},\alpha_{\ell})$. 
Notice that
\[
    C_{f,c} \subseteq 
    \bigcup_{1\leq \ell \leq k} B_\ell +
    \{\alpha_0,\alpha_1,\ldots,\alpha_k\} .
\]
Next, we show that there can be at most $n+1$ critical values in each $B_\ell$. Let there be $1 \leq \ell \leq k$ and some $\beta\in B_\ell$. 
Observe that therefore $\beta \notin \bigcup_{1\leq i\leq n} C_{f_i, p}$. 
For every $i$, let $S_i^\beta \gets \GreedyGSContract(\beta_i, f, p)$ denote the sets produced by running $\AltGreedyGSSPA(\beta, f, p, g)$ in iteration $i$, and denote its result by $S_\beta^*$.  

First, we show that there exists some $\epsilon'>0$, for which running \AltGreedyGSSPA\ on contract $(\beta-\epsilon')$ produces the same sequence $S_1^\beta, \ldots, S_n^\beta$. 
To this end, since $\beta \notin \bigcup_{1\leq i\leq n} C_{f_i, p}$, then for every $i$, $\beta \notin C_{f_i, p}$, so there must exist some $\epsilon_i>0$, for which $\GreedyGSContract((\beta-\epsilon_i), f_i, p)$ returns $S_i$ as well. 
Clearly, for $\epsilon':=\min \{\epsilon_1, \ldots, \epsilon_n \}$, for every $i$, $\GreedyGSContract((\beta-\epsilon'), f_i, p)$ returns $S_i$ as well so the claim holds. 

Next, since $\beta \in C_{f,c}$, there must exist some sufficiently small $\epsilon''>0$, for which running algorithm \AltGreedyGSSPA\ on contracts $\beta$ and $(\beta-\epsilon'')$ yields different results $S_\beta^* \neq S_{\beta-\epsilon''}^*$. 
Therefore, it is clear that for $\epsilon:=\min \{\epsilon', \epsilon''\}$, running \AltGreedyGSSPA\ on contracts $\beta$ and $(\beta-\epsilon)$ produce the same sequence $S_1^\beta, \ldots, S_n^\beta$, but then yield different results $S_\beta^* \neq S_{\beta-\epsilon}^*$.
These different results are chosen in step (\ref{st:alg-alternative-gs-spa-final}).
Let $i,j$ be such that $S_{\beta-\epsilon}^* = S_i^\beta$ and $S_{\beta}^* =S_j^\beta$.
By the monotonicity lemma (Lemma~\ref{lemma:monotonicity-of-critical-values}), we have $f(S_{\beta-\epsilon}^*) < f(S_{\beta}^*)$ so $f(S_i^\beta) < f(S_j^\beta)$.
In addition, by Corollary~\ref{obs:result-of-greedy-gs-f_i-f_j} and the monotonicity of $f$, it follows that $i<j$ and thus  $S_i^\beta \subset S_j^\beta$. 
Overall, $S_{\beta-\epsilon}^* \subset S_{\beta}^*$. 

Now let $B_\ell=\{ \beta_1, \ldots, \beta_r\}$ with $r=|B_\ell|$.
Then $S_{\beta_1}\subset \ldots S_{\beta_r}\subseteq A$, and 
therefore $r\leq n+1$. 
To conclude, $|B_\ell|\leq n+1$ for every $1\leq \ell\leq n $.
We get
\[
    |C_{f,c}| \leq 
    | \bigcup_{1\leq \ell \leq k} B_\ell | + 
    |\{\alpha_0,\alpha_1,\ldots,\alpha_k\} | =
    k(n+1) + k \leq
    \frac{n^2(n+1)(n+2)}{2}.
\]
\end{proof}


\section{Omitted Proofs from Section~\ref{sec:ultra-add}} \label{sec:appendix-ultra-add}

\begin{lemma} \label{lemma:alpha-f-is-ultra}
Let $f:\setfunc$ be an \Ultra\ function, $g:\setfunc$ a symmetric monotone function, and $\alpha \in [0,1]$. 
Define the functions $\alpha f:\setfunc$ and $\alpha f-g:\setfunc$ by
$\alpha f(S) = \alpha \cdot f(S)$ and $(\alpha f-g)(S) = \alpha f(S) - g(S)$ for every set $S$.
Then both $\alpha f$ and $\alpha f-g$ are \Ultra.
\end{lemma}

\begin{proof}
Using the exchange condition (which is equivalent to \Ultra\ by Definition~\ref{def:ultra-new}), it is easy to validate both claims. Let there be any sets $A,B$ with $|A| \leq |B|$, and let there be some $x\in A-B$. From exchange on $f$, it holds that there exists some $y \in B-A$ s.t. $f(A-x+y)+f(B-y+x) \geq f(A) + f(B)$. Multiplying both sides by $\alpha$, we get $\alpha f(A-x+y)+ \alpha f(B-y+x) \geq \alpha f(A) + \alpha f(B)$, thus $\alpha f$ satisfies exchange, so it is \Ultra. Furthermore, from symmetry of $g$, note that $g(A-x+y)=g(A)$ and $g(B-y+x)=g(B)$. Therefore, subtracting $g(A)-g(B)$ from both sides yields  $\alpha f(A-x+y) -g(A-x+y) + \alpha f(B-y+x) -g(B-y+x) \geq \alpha f(A) -g(A)+ \alpha f(B) -g(B)$, thus $\alpha f-g$ satisfies exchange as well, so it is \Ultra.
\end{proof}

\contractdemandqueryultraadd*
\begin{proof}
Let $\Sa^*$ be the set returned by $\GreedyUltraContract(\alpha, f, p)$.
To prove the lemma, we need to show that $\Sa^* \in D_{f,p}(\alpha)$, and $f(\Sa^*) \geq f(S')$ for every $S'\in D_{f,p}(\alpha)$.
The fact that $\Sa^* \in D_{f,p}(\alpha)$ follows by observing that $\GreedyUltraContract(\alpha, f, p)$ is equivalent to  $\GreedyUltraVal(\alpha f, p)$ when the same tie breaking is used, and that if $f$ is \Ultra, then $\alpha f$ is also \Ultra\ (the latter is stated and proved in Lemma~\ref{lemma:alpha-f-is-ultra} in Appendix~\ref{sec:appendix-ultra-add}).
The fact that $f(\Sa^*) \geq f(S')$ for every $S'\in D_{f,p}(\alpha)$ is stated and proved in the next lemma (Lemma~\ref{lemma:ultra-add-demand-maximal-f}).
\end{proof}

\begin{lemma} \label{lemma:ultra-add-demand-maximal-f}
Let $f:\setfunc$ be an \Ultra\ reward function, $p:\setfunc$ an additive cost function, and $\alpha \in [0,1]$ a contract. Let $\Sa^* \gets \GreedyUltraContract(\alpha, f, p)$. 
Then $f(\Sa^*) \ge f(S')$ for every $S' \in D_{f,p}(\alpha)$.
\end{lemma}

\begin{proof}
To see that $f(\Sa^*) \geq f(S')$ for every $S' \in D_{f,p}(\alpha)$, we show that there exists some $\epsilon>0$, for which $\Sa^* \in D_{f,p}(\alpha+\epsilon)$. Thus, from the monotonicity lemma we conclude the claim. 

Let $S_1, \ldots, S_n$ and $x_1, \ldots, x_n$ denote the sequences produced by $\GreedyUltraContract(\alpha, f, p)$.
In step (\ref{st:alg-ultra-demand-query-contract-i}) of Algorithm~\ref{alg:ultra-demand-query-contract}, for every $i$ and for every $x\notin S_{i-1}$ with $x\neq x_i$, one of the following holds.
\begin{enumerate}
    \item \label{en:contract-demand-ultra-no-tiebreak-i}  $\alpha f(x \mid S_{i-1}) - p(x) < \alpha f(x_i \mid S_{i-1}) - p(x_i) $. Or, 
    \item \label{en:contract-demand-ultra-tiebreak-i} $\alpha f(x \mid S_{i-1}) - p(x) = \alpha f(x_i \mid S_{i-1}) - p(x_i) $. Then, by the tie-breaking rule, $f(x_i \mid S_{i-1}) \geq f(x \mid S_{i-1})$.
\end{enumerate}
Similarly, in step (\ref{st:alg-ultra-demand-query-contract-final}) of Algorithm~\ref{alg:ultra-demand-query}, for every $i$ s.t. $S_i\neq \Sa^*$, one of the following holds.
\begin{enumerate}
    \setcounter{enumi}{2}  
    \item \label{en:contract-demand-ultra-no-tiebreak-final} $\alpha f(S_{i}) - p(S_{i}) <  \alpha f(\Sa^*) - p(\Sa^*) $. Or,
    \item \label{en:contract-demand-ultra-tiebreak-final} $\alpha f(S_{i}) - p(S_{i}) = \alpha f(\Sa^*) - p(\Sa^*) $. Thus, by the tie-breaking rule, $f(\Sa^*) \geq f(S_{i})$.
\end{enumerate}
Since there are finitely many strict inequalities of types (\ref{en:contract-demand-ultra-no-tiebreak-i}), (\ref{en:contract-demand-ultra-no-tiebreak-final}), clearly there exists a sufficiently small $\epsilon>0$ such that, for every $i$ and every $x\notin S_{i-1}$ with $x\neq x_i$ that satisfies (\ref{en:contract-demand-ultra-no-tiebreak-i}), it also holds that
\[
(\alpha+\epsilon) f(x \mid S_{i-1}) - p(x) < 
(\alpha+\epsilon) f(x_i \mid S_{i-1}) - p(x_i),
\]
and for every $i$ s.t. $\Sa^*\neq S_i$ that satisfies (\ref{en:contract-demand-ultra-no-tiebreak-final}), it also holds that
\[
(\alpha+\epsilon) f(S_{i}) - p(S_{i}) <  
(\alpha+\epsilon) f(\Sa^*) - p(\Sa^*).
\]
Moreover, every for $i$ and every $x\notin S_{i-1}$ with $x\neq x_i$ that satisfies (\ref{en:contract-demand-ultra-tiebreak-i}), it also holds that
\[
(\alpha+\epsilon) f(x \mid S_{i-1}) - p(x) \leq 
(\alpha+\epsilon) f(x_i \mid S_{i-1}) - p(x_i),
\]
and for every $i$ s.t. $S_i\neq \Sa^*$ that satisfies (\ref{en:contract-demand-ultra-tiebreak-final}), it also holds that
\[
(\alpha+\epsilon) f(S_{i}) - p(S_{i}) \leq  
(\alpha+\epsilon) f(\Sa^*) - p(\Sa^*).
\]
Consequently, executing \GreedyUltraContract\ on contracts $\alpha$ and $(\alpha+\epsilon)$ would follow the same path and return the same result $\Sa^*$, therefore $\Sa^*\in D_{f,p}(\alpha+\epsilon)$. Thus, by the monotonicity lemma (Lemma~\ref{lemma:monotonicity-of-critical-values}), every $S'\in D_{f,p}(\alpha)$ has $f(\Sa^*)\geq f(S')$.
\end{proof}

\genericcostdistinctcosts*
\begin{proof}
Assume that there are two action sets $T_1,T_2$ with $c(T_1)=c(T_2)$. Let $T=T_1 \cap T_2$, then $T'_1:=(T_1\setminus T)$ and $T'_2:=(T_2\setminus T)$ are disjoint.
Since $c(T_1)=c(T_2)$, it follows that for every real value $\alpha$, it holds that $\alpha f(T'_1 \mid A) -c(T'_1)=\alpha f(T'_2 \mid A) -c(T'_2)$, so $\Gamma_{f,c}(T'_1,T'_2)$ must contain all real numbers.
Thus, $c(T_1)\neq c(T_2)$, and in particular, it also holds that $c(T_1) \neq c(\emptyset)$, and that for any two actions $a\neq a'\in A$, $c(a) \neq c(a')$.
\end{proof}


\section{Omitted Proofs from Section~\ref{sec:ft}} \label{sec:appendix-spa-ft}

\demandforspa*
\begin{proof}
Let $S_1, \ldots, S_n$ denote the sequence produced by $\DemandForSPA(\alpha, f, p, g)$, and let $\Sa^*$ denote its result.
To prove the lemma, we need to show that the following two conditions hold: $(i)$ $\Sa^* \in D_{f,p}(\alpha)$, and $(ii)$ $f(\Sa^*) \geq f(S')$ for every $S'\in D_{f,p}(\alpha)$.

To show $(i)$ (i.e., $\Sa^* \in D_{f,c}(\alpha)$), we prove that $\Sa^*$ attains the highest utility among all sets. 
Let $S$ be any set with $k=|S|$ and let $k'=|S_k|$. 
Let $g_i$ be the value of $g$ on a set of size $i$.
By assumption, $S_k \in D^{\leq k}_{f,p}(\alpha)$, and since $|S_k|=k'\leq k$, we have that $g_{k'} \leq g_k$ by monotonicity of $g$. 
Thus,
\[
    \alpha f(S)-p(S)-g(S) = 
    \alpha f(S)-p(S)-g_k \leq 
    \alpha f(S_k)-p(S_k)-g_k \leq
\]\[
    \alpha f(S_k)-p(S_k)-g_{k'} =
    \alpha f(S_k)-p(S_k)-g(S_k) \leq 
    \alpha f(\Sa^*)-p(\Sa^*)-g(\Sa^*).    
\]
The proof of condition $(ii)$ appears in Lemma~\ref{lemma:demand-spa-maximal-f}.
\end{proof}
\begin{lemma} \label{lemma:demand-spa-maximal-f}
Let $f:\setfunc$ be a monotone set function, 
$c=g+p$ a \SPA\ cost function, 
and $\alpha \in [0,1]$ a contract. 
Assume that for every $i\in [n]$, there exists some $\epsilon_i>0$, such that for every $0<\epsilon\leq \epsilon_i$, step (\ref{st:alg-demand-spa-i}) of Algorithm~\ref{alg:demand-spa} generates the same result for $\alpha$ and $(\alpha+\epsilon)$. 
For $\Sa^* \gets \DemandForSPA(\alpha, f, p, g)$, it holds that $f(\Sa^*)\geq f(S')$ for every $S'\in D_{f,c}(\alpha)$.

\end{lemma}
\begin{proof}
Let $S_1, \ldots, S_n$ denote the sequence of sets produced by $\DemandForSPA$($\alpha$, f, p, g), and let $\Sa^*$ denote its result. By assumption, for every $i\in [n]$ there exists some $\epsilon_i>0$, for which running $\DemandForSPA((\alpha+\epsilon), f, p, g)$ gets $S_i$ in iteration $i$, for every $0<\epsilon\leq \epsilon_i$. 
Moreover, there exists some $\epsilon'>0$, for which, in step (\ref{st:alg-demand-spa-final}) of Algorithm~\ref{alg:demand-spa}, for every $i$ and every $S_i$ s.t. $\Sa^*\neq S_i$ that satisfies 
$\alpha f(S_i)-p(S_i)-g(S_i) <
 \alpha f(\Sa^*)-p(\Sa^*)-g(\Sa^*)$, it also holds that
$(\alpha+\epsilon') f(S_i)-p(S_i)-g(S_i) <
 (\alpha+\epsilon') f(\Sa^*)-p(\Sa^*)-g(\Sa^*)$.

Thus, it is easy to verify that for $\epsilon:=\min\{\epsilon_1, \ldots, \epsilon_n, \epsilon'\}$, running \DemandForSPA($(\alpha+\epsilon)$, f, p, g) yields the same sequence $S_1, \ldots, S_n$, and returns the same result $\Sa^*$. 

Consequently, $\Sa^* \in D_{f,c}(\alpha+\epsilon)$, and therefore from the monotonicity lemma (Lemma~\ref{lemma:monotonicity-of-critical-values}), $f(S')\leq f(\Sa^*)$ for any $S'\in D_{f,c}(\alpha)$.
\end{proof}

\ftstructuralinsights*
\begin{proof}
The first statements are straightforward. We next prove statements (\ref{en:f_t-structural-insights-1})-(\ref{en:f_t-structural-insights-4}).

\begin{enumerate}

\item Observe that $\argmax_{S, |S|\leq t} \Big(f(S)-p(S)\Big) = \argmax_{S, |S|\leq t} \Big(f_t(S)-p(S)\Big)$ by the definition of $f_t$. Thus, it suffices to show that $f_t(S^*)-p(S^*) \geq f_t(S)-p(S)$ for every $|S|>t$ as well.
If $|S|>t$, then $f_t(S) = f(\hat{S})$ for some $\hat{S} \in \argmax_{S'\subset S, |S'|=t} {f(S')}$ and $p(\hat{S}) \leq p(S)$ as $\hat{S}\subset S$. Therefore, 
$f_t(S)-p(S) = f(\hat{S}) - p(S) \leq f(\hat{S}) - p(\hat{S}) \leq f(S^*)-p(S^*) = f_t(S^*)-p(S^*)$, where the last inequality is by $S^* \in \argmax_{S, |S|\leq t}{\Big(f(S)-p(S)}\Big)$. 

\item This claim follows from (\ref{en:f_t-structural-insights-1}), since $S^*\in D_f(p)$ and $|S^*|\leq t$ implies that $S^* \in \argmax_{S, |S|\leq t} \Big(f(S)-p(S)\Big)$.

\item We show that $S_t$ has (weakly) higher utility for $f_t$ than any $S$ with $|S|>t$. To this end, let $S$ be an arbitrary set with $|S|>t$, and let $\hat{S} \in \argmax_{S' \subset S, |S'|=t} f(S')$. Thus, $f_t(S)=f(\hat{S})$ by definition of $f_t$. Moreover, as $\hat{S} \subset S$, it holds that $p(\hat{S}) \leq p(S)$.
We get $f_t(S)-p(S) = f(\hat{S}) - p(S) \leq f(\hat{S}) -p(\hat{S}) \leq f(S_t)- p(S_t) =f_t(S_t)-p(S_t)$, where the last inequality holds since $|\hat{S}|=t$, implying that $S_t$ has at least the same utility for $f,p$ than $\hat{S}$. The last equality follows by definition of $f_t$. 

\item This claim clearly holds by the definition of $f_t$. For any set $S$ with $|S| \leq t$, it holds that 
$f(S)-p(S)=f_t(S)-p(S)\leq f_t(S^*)-p(S^*) = f(S^*)-p(S^*)$.
\end{enumerate}
\end{proof}


\section{Omitted Proofs from Section~\ref{sec:ultra-spa}} \label{sec:appendix-ultra-spa}


\begin{lemma} \label{lemma:ultra-spa-maximal-f}
Let $f:\setfunc$ be an \Ultra\ reward function, $c=g+p$ an \SPA\ cost function, and $\alpha \in [0,1]$ a contract. Let $\Sa^* \gets \GreedyUltraSPA(\alpha, f, p, g)$. Then $f(\Sa^*) \ge f(S')$ for every $S' \in D_{f,c}(\alpha)$.
\end{lemma}

\begin{proof}
To see that $f(\Sa^*) \geq f(S')$ for every $S'\in D_{f,c}(\alpha)$, we show that there exists some $\epsilon>0$, for which $\Sa^* \in D_{f,c}(\alpha+\epsilon)$. Thus, from the monotonicity lemma we conclude the claim. To this end, let $S_1, \ldots, S_n$ and $x_1, \ldots, x_n$ denote the sequences produced by running $\GreedyUltraSPA(\alpha, f, p, g)$. In step (\ref{st:alg-ultra-spa-demand-query-contract-i}) of Algorithm~\ref{alg:ultra-spa-demand-query-contract}, for every $i$ and for every $x\notin S_{i-1}$ with $x\neq x_i$, one of the following holds.
\begin{enumerate}
    \item \label{en:contract-demand-ultra-spa-no-tiebreak-i}  $\alpha f(x \mid S_{i-1}) - p(x)  -  g(x \mid S_{i-1}) < \alpha f(x_i \mid S_{i-1}) - p(x_i)-g(x_i \mid S_{i-1})  $, or, 
    \item \label{en:contract-demand-ultra-spa-tiebreak-i}  $\alpha f(x \mid S_{i-1}) - p(x)  -  g(x \mid S_{i-1}) = \alpha f(x_i \mid S_{i-1})- p(x_i) -g(x_i \mid S_{i-1}) $. Then, by the tie-breaking rule, $f(x_i \mid S_{i-1}) \geq f(x \mid S_{i-1})$.
\end{enumerate}
Similarly, in step (\ref{st:alg-ultra-spa-demand-query-contract-final}) of Algorithm~\ref{alg:ultra-spa-demand-query-contract}, for every $i$ s.t. $S_i\neq \Sa^*$, one of the following holds.
\begin{enumerate}
    \setcounter{enumi}{2}  
    \item \label{en:contract-demand-ultra-spa-no-tiebreak-final} $\alpha f(\Sa^*)- p(\Sa^*)  -g(\Sa^*) <  \alpha f(S_{i}) - p(S_{i}) -g(S_i)$, or,
    \item \label{en:contract-demand-ultra-spa-tiebreak-final} $\alpha f(\Sa^*) - p(\Sa^*) - g(\Sa^*) = \alpha f(S_{i}) - p(S_{i}) - g(S_{i})$. Therefore, by the tie-breaking rule $f(\Sa^*) \geq f(S_{i})$.
\end{enumerate}
Since there are finitely many strict inequalities of types (\ref{en:contract-demand-ultra-spa-no-tiebreak-i}), (\ref{en:contract-demand-ultra-spa-no-tiebreak-final}), clearly there exists a sufficiently small $\epsilon>0$ such that, for every $i$ and every $x\notin S_{i-1}$ with $x\neq x_i$ that satisfies (\ref{en:contract-demand-ultra-spa-no-tiebreak-i}), it also holds that 
\[
(\alpha+\epsilon) f(x \mid S_{i-1}) - p(x) -g(x \mid S_{i-1}) < 
(\alpha+\epsilon) f(x_i \mid S_{i-1})  - p(x_i) -g(x_i \mid S_{i-1}),
\]
and for every $i$ s.t. $\Sa^*\neq S_i$ that satisfies (\ref{en:contract-demand-ultra-spa-no-tiebreak-final}), it also holds that
\[
(\alpha+\epsilon) f(\Sa^*) - p(\Sa^*)  - g(\Sa^*)<  
(\alpha+\epsilon) f(S_{i}) - p(S_{i})  - g(S_{i}).
\]
Furthermore, for every $i$ and every $x\notin S_{i-1}$ with $x\neq x_i$ that satisfies (\ref{en:contract-demand-ultra-spa-tiebreak-i}), it also holds that
\[
(\alpha+\epsilon) f(x \mid S_{i-1}) - p(x) -g(x \mid S_{i-1}) \leq
(\alpha+\epsilon) f(x_i \mid S_{i-1}) - p(x_i)-g(x_i \mid S_{i-1}),
\]
and for every $i$ s.t. $S_i\neq \Sa^*$ that satisfies (\ref{en:contract-demand-ultra-spa-tiebreak-final}), it also holds that 
\[
(\alpha+\epsilon) f(\Sa^*)-p(\Sa^*)  - g(\Sa^*)\leq  
(\alpha+\epsilon) f(S_{i})- p(S_{i}) - g(S_{i}).
\]
Consequently, executing \GreedyUltraSPA\ on contracts $\alpha$ and $(\alpha+\epsilon)$ would follow the same path and return the same result  $\Sa^*$, therefore  $\Sa^*\in D_{f,c}(\alpha+\epsilon)$. Thus, by the monotonicity lemma (Lemma~\ref{lemma:monotonicity-of-critical-values}), every $S'\in D_{f,c}(\alpha)$ has $f(\Sa^*)\geq f(S')$.
\end{proof}

\begin{lemma} \label{lemma:greedy-up-to-t-alpha-plus-epsilon}
    Let an \Ultra\ reward function $f:\setfunc$, an additive cost function $c:\setfunc$, and some $t\in [n]$.
    For every contract $\alpha\in[0,1]$, there exists some $\epsilon'>0$, such that for every $0<\epsilon\leq \epsilon'$, $\GreedyUpToT(\alpha, t, f, c)$ and $\GreedyUpToT((\alpha+\epsilon), t, f, c)$ returns the same result.
\end{lemma}
\begin{proof}
The claim follows from the structure of \GreedyUpToT. 
Observe that in step (\ref{st:alg-greedy-up-to-t-i}) of Algorithm~\ref{alg:greedy-up-to-t}, for every $i\leq t$, and every $x\notin S_{i-1}$ with $x\neq x_i$, one of the following holds: 
\begin{enumerate}
\item \label{en:f_t_no_tiebreak_i} $\alpha f(x \mid S_{i-1})-p(x) < \alpha f(x_i \mid S_{i-1})-p(x_i)$. Or,
\item \label{en:f_t_tiebreak_i} $\alpha f(x \mid S_{i-1})-p(x) = \alpha f(x_i \mid S_{i-1})-p(x_i)$. Then by the tie-breaking rule, $f(x \mid S_{i-1})\leq f(x_i \mid S_{i-1})$.
\end{enumerate}
Similarly, in the final step (\ref{st:alg-greedy-up-to-t-final}) of the algorithm, for any $i\leq t$ with $S_i\neq \Sa^*$, one of the following holds: 
\begin{enumerate}[resume]
\item \label{en:f_t_no_tiebreak_final} $\alpha f(S_i)-p(S_i) < \alpha f(\Sa^*)-p(\Sa^*)$. Or,
\item \label{en:f_t_tiebreak_final} $\alpha f(S_i)-p(S_i) = \alpha f(\Sa^*)-p(\Sa^*)$. Therefore, by the tie-breaking rule, $f(S_i)\leq f(\Sa^*)$.
\end{enumerate}
Since there are finitely many inequalities of types (\ref{en:f_t_no_tiebreak_i}), (\ref{en:f_t_no_tiebreak_final}), there has to exists some sufficiently small $\epsilon'>0$, such that for every $0<\epsilon\leq \epsilon'$, the following holds: 
for every $i\leq t$ and every $x\notin S_{i-1}$ with $x\neq x_i$ that satisfies (\ref{en:f_t_no_tiebreak_i}), it also holds that 
\[
    (\alpha+\epsilon) f(x \mid S_{i-1})-p(x) < 
    (\alpha+\epsilon) f(x_i \mid S_{i-1})-p(x_i),
\]
and for every $i\leq t$ s.t. $S_i\neq \Sa^*$ that satisfies (\ref{en:f_t_no_tiebreak_final}), it also holds that
\[
    (\alpha+\epsilon) f(S_i)-p(S_i) < 
    (\alpha+\epsilon) f(\Sa^*)-p(\Sa^*).
\]
Moreover, for every $i\leq t$ and every $x\notin S_{i-1}$ with $x\neq x_i$ that has (\ref{en:f_t_tiebreak_i}), 
it also holds that 
\[
(\alpha+\epsilon) f(x \mid S_{i-1})-p(x) \leq 
(\alpha+\epsilon) f(x_i \mid S_{i-1})-p(x_i),
\]
and for every $i\leq t$ s.t. $S_i\neq \Sa^*$ that satisfies (\ref{en:f_t_tiebreak_final}), it also holds that
\[
(\alpha+\epsilon) f(S_i)-p(S_i) \leq 
(\alpha+\epsilon) f(\Sa^*)-p(\Sa^*).
\]
Consequently, executing \GreedyUpToT\ on contracts $\alpha$ and $(\alpha+\epsilon)$ would follow the same path and return the same result $\Sa^*$. 
\end{proof}

\begin{lemma} \label{lemma:greedy-up-to-t-maximal-f}
Consider an \Ultra\ function $f:\setfunc$, and $f_t$ that is the $t$-truncation of $f$ for some $t\in[n]$ (See Def.~\ref{def:f_t}). Consider also an additive cost $p:\setfunc$ and a contract $\alpha\in [0,1]$. Let $\Sa^*\gets\GreedyUpToT(\alpha, t, f, p)$. Then, $f_t(\Sa^*)\geq f_t(S')$ for every $S'\in D_{f_t,p}(\alpha)$.
\end{lemma}

\begin{proof}
To see that $f_t(\Sa^*)\geq f_t(S')$ for every $S'\in D_{f_t,p}(\alpha)$, observe that from Lemma~\ref{lemma:greedy-up-to-t-alpha-plus-epsilon}, there exists some $\epsilon>0$ for which $\GreedyUpToT((\alpha+\epsilon), f, p)$ returns $\Sa^*$. Therefore, it holds that $\Sa^* \in D_{f,p}(\alpha+\epsilon)$ as well. As a result, by the monotonicity lemma (Lemma~\ref{lemma:monotonicity-of-critical-values}), any $S'\in D_{f_t,p}(\alpha)$ has $f_t(S') \leq f_t(\Sa^*)$.  
\end{proof}

\greedyuptotSisubsetSj*
\begin{proof}
Let $S_1^{(i)}, \ldots S_i^{(i)}$ denote the sequences produced by $\GreedyUpToT(\alpha, i,f,p)$ and let $S^*_{(i)}$ denote its result. It is immediate from the definition of $f_i$ and $f_j$ that up to iteration $i$, all the computations made by the algorithm for $f_i$ and $f_j$ are the same. Having said that, let $S_1^{(i)}, \ldots S_i^{(i)}, S_{i+1}^{(j)}, \ldots, S_{j}^{(j)}$ denote the sequences produced by $\GreedyUpToT(\alpha, j,f,p)$, and let $S^*_{(j)}$ denote its result. 

Let $i^*\leq i$ be such that $S_{(i)}^*=S_{i^*}^{(i)}$, and $j^*\leq j$ be such that $S_{(j)}^*=S_{j^*}^{(j)}$ (these are the indices of sets that are returned by the algorithm on $f_i$ and on $f_j$, respectively).
Notice that by the definition of the algorithm, $S_i^{(i)}\subseteq S_{j'}^{(j)}$ for every $j'>i$. Thus, it suffices to prove that $i^*\leq j^*$ and the claim will follow. 
Assume towards contradiction that $j^*<i^*\leq i$. It then holds that $S_{(j)}^*=S_{j^*}^{(j)}=S_{j^*}^{(i)}$. Moreover, it holds that
\begin{enumerate} [itemsep=0pt, topsep=0pt, leftmargin=15pt]
    \item By $\GreedyUpToT(\alpha, i,f,p)$, $S_{i^*}^{(i)}$ has maximal utility out of $S_1^{(i)}, \ldots S_i^{(i)}$. In particular, it has a weakly higher utility then $S_{j^*}^{(i)}$ (as $j^*< i$), so $\alpha f(S_{i^*}^{(i)})-p(S_{i^*}^{(i)}) \geq \alpha f(S_{j^*}^{(i)})-p(S_{j^*}^{(i)})$.
    \item By $\GreedyUpToT(\alpha, j,f,p)$, $S_{j^*}^{(i)}$ has maximal utility out of $S_1^{(j)}, \ldots S_i^{(j)}$. In particular, it has a weakly higher utility then $S_{i^*}^{(i)}$ (as $i^*< j$), so $\alpha f(S_{i^*}^{(i)})-p(S_{i^*}^{(i)}) \geq \alpha f(S_{j^*}^{(i)})-p(S_{j^*}^{(i)})$.
\end{enumerate}
Therefore, $S_{i^*}^{(i)}$ and $S_{j^*}^{(i)}$ must have equal utilities. But by the consistent tie breaking, it is not possible that the final outcome was different, which is a contradiction. Altogether, $i^*\leq j^*$, and therefore $S_{(i)}^*\subseteq S_{(j)}^*$. 
\end{proof}

\end{document}